\documentclass[aps, prd, showpacs,floatfix, superscriptaddress]{revtex4-2}
\usepackage{amsmath,amssymb,amsfonts,mathtools}
\usepackage{float}
\usepackage{graphicx}
\usepackage{xcolor}
\usepackage{tensor}
\usepackage{lineno}
\usepackage[colorlinks,linkcolor=blue,citecolor=blue,urlcolor=blue]{hyperref}
\usepackage{appendix}
\usepackage{comment}

\usepackage[T1]{fontenc}
\usepackage{cancel}
\usepackage{soul}
\usepackage{ulem}

\newcommand{\D}{\text{\sout{$\Delta$}}}

\newcommand{\cll}{\text{\sout{$<$}}}
\newcommand{\crr}{\text{\sout{$>$}}}
\newcommand{\h}{\hspace{1.8mm}}
\usepackage{braket}

\begin{document}

\title{Second-order spin hydrodynamics from Zubarev's nonequilibrium statistical operator formalism}

\date{ \today }

\author{Abhishek Tiwari\footnote{abhi7phy@gmail.com}}
\author{Binoy Krishna Patra\footnote{binoy@ph.iitr.ac.in}}

\affiliation{Department of Physics, Indian Institute of Technology Roorkee, Roorkee 247667, India}

%
\begin{abstract} 
Using the Zubarev's nonequilibrium statistical operator formalism, we derive the second-order expression for the dissipative tensors in relativistic spin hydrodynamics, {\em viz.}  rotational stress tensor ($\tau_{\mu\nu}$), boost heat vector ($q_\mu$), shear stress tensor ($\pi_{\mu\nu}$), and bulk viscous pressure ($\Pi$). The first two ($\tau_{\mu\nu}$ and $q_\mu$) emerge due to the inclusion of the antisymmetric part in the energy-momentum tensor, which, in turn, governs the conservation of spin angular momentum ($\Sigma^{\alpha\mu\nu}$). As a result, new thermodynamic forces, generated due to the antisymmetric part of $T_{\mu \nu}$, contain the spin chemical potential. In this work, we have also taken the spin density ($S^{\mu \nu}$) as an independent thermodynamic variable, in addition to the energy density and particle density, thereby resulting in two novel transport coefficients given by the correlation between spin density tensor and rotational stress tensor and vice versa. Additionally, the newly found terms in $\pi_{\mu\nu}$ and $\Pi$ are the artifacts of the new thermodynamic forces that arise due to the antisymmetric part of $T^{\mu \nu}$. Finally, we have derived the evolution equations for the aforesaid tensors: $\tau_{\mu\nu}$, $q_\mu$, $\pi_{\mu\nu}$, and $\Pi$.
\end{abstract}
%

\maketitle


\section{Introduction}\label{introdunction}

Ever since the experiments~\cite{STAR:2017ckg, 
	STAR:2018gyt,STAR:2019erd} measuring the spin polarization of 
$\Lambda$ hyperons, there has been a growing interest in studying 
spin-dependent observables (see~\cite{BecattiniSprhic, NiidaPphic} for a recent review on spin polarization). Also, the study of spin transport 
phenomena in condensed matter systems, known as 
spintronics~\cite{Maekawa}, has become a prominent research area. 
In addition to this, despite the success of relativistic 
hydrodynamics in various areas, it has been unable to provide a 
comprehensive explanation for the experimental data on the differential spin polarization of $\Lambda$ hyperons. This problem is known as the ``spin sign problem''~\cite{STAR:2018gyt,STAR:2019erd}~(see~\cite{Becattini:2020ngo, Florkowski:2018fap} for review). An effective way to address these findings is by 
incorporating the spin degree of freedom. Significant 
efforts have been made to explain the spin polarization in heavy ion collisions~\cite{Karpenko:2016jyx, Fang:2016vpj, Xie:2017upb, Sun:2017xhx, Florkowski:2019qdp, Liu:2020dxg, Shi:2020htn2, Pang:2021, Alzhrani:2022, Buzzegoli:2024ghi, Palermo:2024, Weickgenannt:2024, Fang:2023, Bhadury:2023vjx, Florkowski:2025abc, Florkowski:2025def, Dey:2025ail} and to develop relativistic hydrodynamics that includes 
spin, known as spin-hydrodynamics (see~\cite{Huang:2024ffg} for a recent review on spin hydrodynamics).

The intriguing feature of spin hydrodynamics is captured in recent studies~\cite{Hattori:2019lfp, Fukushima:2020ucl}, where two new transport coefficients arise, reflecting the physical effect of spin. These new transport coefficients, namely boost heat conductivity and rotational viscosity, originate at first-order and tell the slow dynamics of spinful fluid systems~\cite{Hattori:2019lfp}. These transport coefficients emerge due to the inclusion of new dissipative tensors via the antisymmetric part of the energy-momentum tensor.

Transport coefficients are crucial, as they govern the evolution of dissipative tensors present in the medium. Each dissipative tensor evolves over time, capturing particular physical aspects of the medium. At the second order, new transport coefficients arise, and the effect of spin can be seen in the traditional dissipative tensors, such as bulk viscous pressure and shear stress tensor. Moreover, the first-order theory is acausal, and in order to restore causality, a second-order theory is needed. Also, a numerical implementation of causal spin hydrodynamic theory may give information about the spin sign problem in heavy-ion collisions. These results strongly motivate the development of a theoretical framework that describes spin transport in relativistic plasma.

Various approaches have been 
proposed to incorporate spin into hydrodynamics, including the 
quantum kinetic theory approach~\cite{Florkowski:2017ruc, 
	Gao:2019znl, Hattori:2019ahi, Li:2019qkf, Yang:2020hri, 
	Weickgenannt:2020aaf, Liu:2020flb, Bhadury:2020puc, 
	Shi:2020htn}, approaches based on the second law of 
thermodynamics~\cite{Hattori:2019lfp, Fukushima:2020ucl, 
	Li:2020eon, She:2021lhe, Hongo:2021ona}, the effective 
Lagrangian approach~\cite{Montenegro:2017lvf, Montenegro:2017rbu, 
	Montenegro:2018bcf, Montenegro:2020paq}, the equilibrium 
partition function approach~\cite{Jensen:2012jh, 
	Gallegos:2021bzp, Gallegos:2022jow}, entropy current 
analysis~\cite{Becattini:2023ouz, Biswas:2023qsw}, and the 
quantum statistical density operator~\cite{Becattini:2007nd, 
	Becattini:2009wh, Becattini:2012pp, 
	Becattini:2018duy,Hu:2021lnx}.

This paper employs Zubarev's nonequilibrium statistical operator 
(NESO)~\cite{zubarev1979derivation, van1982maximum} to develop a 
framework for second-order spin hydrodynamics. Zubarev's NESO 
method describes nonequilibrium processes and incorporates 
irreversibility in the description. The introduction of 
irreversibility is achieved by incorporating an infinitesimal 
source term into the Liouville equation. The NESO is an extension 
of the Gibbs canonical ensemble, {\em i.e.} the Gibbs relation 
remains locally valid. By utilizing this formalism, we can 
distinctly separate the equilibrium and nonequilibrium 
components. We have derived the second-order expression for 
rotational stress tensor ($\tau_{\mu \nu}$), boost heat vector 
($q_\mu$), shear stress tensor ($\pi_{\mu \nu}$) and bulk viscous 
pressure ($\Pi$) by incorporating the antisymmetric part in the 
energy-momentum tensor, in addition to the symmetric part. 
Moreover, using this method, the transport coefficients of the system can be 
naturally obtained in the form of a Kubo-type formula.

This paper is structured as follows: Section II deals with the 
equations of motion for the energy-momentum tensor, charge, and 
angular momentum. We also discuss the entropy production using 
the extended thermodynamic relation involving the spin chemical 
potential and spin density, in addition to the conventional 
thermodynamic variables. In Section III, we introduce the 
nonequilibrium statistical operator (NESO) and calculate the 
nonequilibrium part of the nonequilibrium statistical operator in 
terms of hydrodynamic gradient expansion. In Section IV, we 
revisit the first-order spin hydrodynamics and derive the 
transport coefficients that emerge in first-order spin 
hydrodynamics. In Section V, we derive the second-order 
expressions and the evolution equations for the rotational stress 
tensor ($\tau_{\mu\nu}$), boost heat vector ($q_\mu$), shear 
stress tensor ($\pi_{\mu\nu}$), and bulk viscous pressure ($\Pi$). 
Finally, we conclude with a summary and discussion in Section VI.


\section{Review of Relativistic Spin Hydrodynamics}
Throughout this paper, we employ the metric tensor 
$g_{\mu\nu}={\rm diag}(1,-1,-1,-1)$. The projection operator is 
defined as $\Delta_{\mu\nu}=g_{\mu\nu}-u_\mu u_\nu$, with the 
normalization condition for the fluid velocity given by $u_\mu 
u^\mu=1$. Given our focus on investigating the spin effect, we 
adopt the Landau frame where $h_\mu=0$. In addition, the 
projectors and other shorthand notations used in this paper are
\begin{align}
	\Delta_{\mu\nu}^{\lambda\delta}&=\frac{1}{2}(\Delta^\lambda_\mu\Delta^\delta_\nu+\Delta^\lambda_\nu\Delta^\delta_\mu)-\frac{1}{3}\Delta^{\lambda\delta}\Delta_{\mu\nu},~~~ \D_{\mu\nu}^{\lambda\delta}=\frac{1}{2}(\Delta^\lambda_\mu\Delta^\delta_\nu-\Delta^\lambda_\nu\Delta^\delta_\mu),\label{shorthand1}\\
	A^{(\mu\nu)} &= \frac{1}{2}(A^{\mu\nu}+A^{\nu\mu}), ~~~A^{[\mu\nu]} = \frac{1}{2}(A^{\mu\nu}-A^{\nu\mu}),\label{shorthand2}\\
	A^{<\mu\nu>} &= \Delta_{\lambda\delta}^{\mu\nu} A^{\lambda\delta}, ~~~ A^{\cll\mu\nu\crr} = \D_{\lambda\delta}^{\mu\nu} A^{\lambda\delta}.\label{shorthand3}
\end{align}

\subsection{Equation of motions}

To start with the derivation of hydrodynamic equations of motion, 
we assume that the energy-momentum tensor $T^{\mu\nu}$, charge 
current $N^\mu$ and total angular momentum $J^{\mu\alpha\beta}$ 
are conserved for the system of interest~\cite{Hattori:2019lfp,  
	She:2021lhe}. The conservation laws are given by
\begin{align}
	\partial_\mu T^{\mu\nu}&=0,  \label{Tconserve}\\
	\partial_\mu N^\mu&=0,  \label{Nconserve}\\
	\partial_\mu J^{\mu\alpha\beta}&=0. \label{Jconserve}
\end{align}
The orbital and the spin part contribute to the total angular 
momentum. Microscopically, it is known that 
\begin{equation}
	J^{\mu\alpha\beta} = x^\alpha T^{\mu\beta}-x^\beta T^{\mu\alpha} + \Sigma^{\mu\alpha\beta} ,
\end{equation}
where $\Sigma^{\mu\alpha\beta}$ is the intrinsic spin degree of 
freedom. $J^{\mu\alpha\beta}$ and $\Sigma^{\mu\alpha\beta}$ are 
antisymmetric in the last two indices. The conservation law for 
$J^{\mu\alpha\beta}$ restrict $\Sigma^{\mu\alpha\beta}$ to follow 
\begin{equation}\label{sigmaconserve}
	\partial_\mu \Sigma^{\mu\alpha\beta} =- \Big(T^{\alpha\beta}-T^{\beta \alpha}\Big).
\end{equation}
The next step involves writing a systematic gradient expansion 
for  $T^{\mu\nu}$, $N^\mu$, and $\Sigma^{\mu\alpha\beta}$. A general tensor decomposition is given by
\begin{align}
	T^{\mu\nu}&=\varepsilon u^\mu u^\nu -(p+\Pi)\Delta^{\mu\nu}+h^\mu u^\nu+h^\nu u^\mu +\pi^{\mu\nu} +q^\mu u^\nu-q^\nu u^\mu + \tau^{\mu\nu},\label{Tmn}\\
	N^\mu &= nu^\mu + j^\mu,\label{Nu}\\
	\Sigma^{\mu\alpha\beta}&= u^\mu S^{\alpha\beta} ,\label{sigmauab}
\end{align}
where $\varepsilon$ is the energy density, $n$ is the particle 
density, $S^{\alpha\beta}$ is the spin density, and $p$ is the 
equilibrium pressure. We have adopted the phenomenological form of the spin tensor, given by $\Sigma^{\mu\alpha\beta}= u^\mu S^{\alpha\beta}$. This choice is motivated by its intuitive physical interpretation: it represents the flow of spin density, where the spin density $S^{\alpha\beta}$ is defined in the local rest frame of the fluid. Moreover, this form is consistent with the structure of the global equilibrium solution of thermal vorticity~\cite{amaresh, Florkowski:2018fap}. Furthermore, it is important to note that 
$u^\mu$, $\varepsilon$, $p$, $n$, $\Delta_{\mu\nu}$ are zeroth 
order term $\mathcal{O}(\partial^0)$ in gradient expansion,
whereas, the dissipative terms $\Pi$, $h^\mu$, $q^\mu$, $j^\mu$, 
$\pi^{\mu\nu}$, $\tau^{\mu\nu}$ in general, contain the gradient correction up to all orders, and they satisfy the 
conditions: $u_\mu h^\mu=0$, $u_\mu q^\mu=0$, $u_\mu j^\mu=0$, $u_\mu \pi^{\mu\nu}=0$, $u_\mu \tau^{\mu\nu}=0$, $\pi^{\mu\nu}=\pi^{\nu\mu}$, $\pi_\mu^\mu=0$, and $\tau^{\mu\nu}=-\tau^{\nu\mu}$.

To calculate the equations of motion, first, we substitute the 
gradient expansion~\eqref{Tmn} in conservation 
equation~\eqref{Tconserve} and take the projection 
of~\eqref{Tconserve} along $u_\nu$ and $\Delta_\nu^\alpha$. We obtain
\begin{align}
		& D\varepsilon +(p+\Pi+\varepsilon)\theta + \partial_\mu h^\mu + u_\mu Dh^\mu + u_\nu\partial_\mu\pi^{\mu\nu}+ \partial_\mu q^\mu - u_\mu Dq^\mu  +u_\nu\partial_\mu\tau^{\mu\nu} =0,\label{Depsilon}\\
	& hDu^\alpha+ \Pi Du^\alpha -\nabla^\alpha(p+\Pi) + h^\alpha\theta+h^\mu\partial_\mu u^\alpha + \Delta^\alpha_\nu Dh^\nu + \Delta^\alpha_\nu\partial_\mu\pi^{\mu\nu}-q^\alpha\theta+ q^\mu\partial_\mu u^\alpha - \Delta^\alpha_\nu Dq^\nu + \Delta^\alpha_\nu\partial_\mu\tau^{\mu\nu} =0.\label{Du}
\end{align}
Here, we defined $\theta \equiv\partial_\mu u^\mu$ and $ D\equiv u^\mu 
\partial_\mu$. Further, we use~\eqref{Nu} and~\eqref{sigmauab} in~\eqref{Nconserve} 
and~\eqref{sigmaconserve}, respectively. Thus we obtain
\begin{align}
	& Dn +n\theta +\partial_\mu j^\mu =0,\label{Dn}\\
	& DS^{\alpha\beta} +S^{\alpha\beta} \theta +2(q^\alpha u^\beta-q^\beta u^\alpha + \tau^{\alpha\beta}) =0.\label{Dsigma}
\end{align}


\subsection{Entropy Production}
In order to incorporate the spin into the description, we treat 
$\Sigma^{\mu\alpha\beta}$ to be an independent hydrodynamic 
field, the thermodynamic relations near local thermodynamics 
equilibrium get modified as~\cite{Hattori:2019lfp}
\begin{align}
	Tds &= d\varepsilon - \mu dn -\frac{1}{2} \omega_{\mu\nu}dS^{\mu\nu}  ~~~~~\text{(first law),} \label{tds}\\
	dp &= sdT + nd\mu  + \frac{1}{2} S^{\mu\nu} d\omega_{\mu\nu} ~~~~~\text{(Gibbs-Duhem),} \label{dp}\\
	\varepsilon+p &= Ts +\mu n+ \frac{1}{2}\omega_{\mu\nu} S^{\mu\nu}  ~~~~~\text{(Euler's relation).} \label{h}
\end{align}
Here, $s$ is the entropy density, $T$ is the temperature, and $\mu$ is 
the chemical potential conjugate to the particle density $n$. 
Similarly, $\omega_{\mu\nu}$ is the spin chemical potential 
conjugate to spin density $S^{\mu \nu}$.

The above thermodynamic relations are essential to write the 
entropy current. However, it has been shown that the form of entropy 
current can be derived from the first principle, using the quantum 
statistical density 
operator~\cite{becattini2019extensivity,BECATTINI2024138533}. The 
form of entropy current, treating spin tensor 
$\Sigma^{\mu\alpha\beta}$ as an independent hydrodynamic field, 
is given by
\begin{equation}
	S^\mu = p\beta^\mu + \beta_\nu T^{\mu\nu} -\alpha N^\mu -\frac{1}{2}\Omega_{\alpha\beta}\Sigma^{\mu\alpha\beta}.\label{encurrent}
\end{equation}
The parameters $\beta^\mu$, $\alpha$ and 
$\Omega_{\alpha \beta}$ are defined by
\begin{equation}
	\beta^\mu=\beta u^\mu, ~~~ \alpha=\beta\mu,~~~ \Omega_{\alpha\beta}=\beta \omega_{\alpha\beta}. \label{connection}
\end{equation}
where, $\beta$ is the inverse temperature, and $u^\mu$ is the fluid 
four velocity. 

According to the second law of thermodynamics, the divergence of 
entropy current, i.e., the entropy production rate, should be 
positive for a dissipative system; therefore,
\begin{equation}\label{enpro}
	\partial_\mu S^\mu \geq 0 .
\end{equation}
Using the gradient expansions for $T^{\mu\nu}$, $N^\mu$ and 
$\Sigma^{\mu\alpha\beta}$ from~\eqref{Tmn},~\eqref{Nu} 
and~\eqref{sigmauab}, respectively, the divergence of entropy 
current is given by
\begin{align}\label{endiv}
	\partial_\mu S^\mu &= -\Pi \beta \theta + \beta h^\mu (\beta^{-1}\partial_\mu \beta+ Du_\mu) -j^\mu \partial_\mu\alpha+\beta q^\mu M_\mu + \beta \pi^{\mu\nu}\sigma_{\mu\nu} +\beta\tau^{\mu\nu}\xi_{\mu\nu},
\end{align}
here,  $M_\mu\equiv (\beta^{-1}\partial_\mu\beta- 
Du_\mu+2\beta^{-1}\Omega_{\mu\nu}u^\nu)$, 
$\sigma_{\mu\nu}\equiv\Delta_{\mu\nu}^{\lambda\delta} 
\partial_\lambda u_\delta$, and 
$\xi_{\mu\nu}\equiv\D_{\mu\nu}^{\lambda\delta}\big( 
\partial_\lambda 
u_\delta+\beta^{-1}\Omega_{\lambda\delta}\big)$. The $\D_{\mu\nu}^{\lambda\delta}
\partial_\lambda 
u_\delta$ term in $\xi_{\mu\nu}$ is known as the vorticity tensor ($\overline{\omega}_{\mu\nu}$).

We would like to emphasize that the hydrodynamic gradient orderings assigned to $\Omega_{\alpha\beta}$ and the spin density $S^{\alpha\beta}$ vary across the literature. For instance, in \cite{She:2021lhe}, both $\Omega_{\alpha\beta}$ and $S^{\alpha\beta}$ are treated as leading-order quantities, $i.e.$, $\mathcal{O}(\partial^0)$. In contrast, other works such as~\cite{Hongo:2021ona, Dong} adopt the convention that both are of order $\mathcal{O}(\partial)$. In the present work, we assign $\Omega_{\alpha\beta} \sim \mathcal{O}(\partial^1)$ and $S^{\alpha\beta} \sim \mathcal{O}(\partial^0)$ within the framework of the hydrodynamic gradient expansion. The choice of $S^{\alpha\beta}$ as a zeroth-order quantity is motivated by analogy with the energy density $\varepsilon$ and particle density $n$, both of which are considered to be of $\mathcal{O}(\partial^0)$. On the other hand, it is well-established that in global equilibrium, the spin chemical potential $\omega_{\alpha\beta}$ can be expressed in terms of the thermal vorticity, which is of $\mathcal{O}(\partial)$. Consequently, the natural ordering for both $\omega_{\alpha\beta}$ or $\Omega_{\alpha\beta}$ is $\mathcal{O}(\partial)$. This physically motivated ordering scheme is widely employed in the literature~\cite{Hattori:2019lfp, Fukushima:2020ucl, Biswas:2023qsw, Hu:2021lnx}. A natural question that arises is how to relate $S^{\alpha\beta}$ and $\Omega^{\alpha\beta}$ when they are assigned different orders. An illustrative example is provided in \cite{boost}. However, if one aims to describe a system with strong vorticity, the ordering scheme $\Omega_{\alpha\beta} \sim \mathcal{O}(\partial^1)$ and $S^{\alpha\beta} \sim \mathcal{O}(\partial^0)$ becomes inadequate. In such cases, it is necessary to treat the vorticity as a leading-order term~\cite{Huang:2024ffg}.

Having clarified that point, we now return to the calculations. Employing the ordering scheme $\Omega_{\alpha\beta} \sim \mathcal{O}(\partial^1)$ and $S^{\alpha\beta} \sim \mathcal{O}(\partial^0)$ and enforcing the 
condition~\eqref{enpro} on~\eqref{endiv}, we obtain the relation 
between the dissipative tensors and hydrodynamic forces up to
first order:
\begin{align}\label{transport}
&	\Pi = -\zeta \theta,\nonumber\\
&\pi^{\mu\nu} = 2 \eta \sigma^{\mu\nu},\nonumber\\
&q^\mu = -\lambda M^\mu,\nonumber\\
&\tau^{\mu\nu} = 2 \gamma \xi^{\mu\nu}.
\end{align}
The transport coefficients $\zeta$ and $\eta$ are known as bulk 
viscosity and shear viscosity, respectively. The other two transport 
coefficients, $\lambda$ and $\gamma$, are new in spin 
hydrodynamics. They are known as boost heat conductivity and 
rotational viscosity, respectively~\cite{Hattori:2019lfp}.


\section{Nonequilibrium statistical operator}
This method starts by defining a nonequilibrium statistical 
operator (NESO) $\hat{\rho}(t)$ for a system that is in a 
hydrodynamic regime~\cite{zubarev1979derivation,van1982maximum}. 
For a system containing spin degrees of freedom, the operator, 
$\hat{\rho}(t)$ is defined as~\cite{HARUTYUNYAN2022168755, 
	Hu:2021lnx}
\begin{equation}
	\hat{\rho}(t) = Q^{-1}\exp\Big[ -\int d^3x \hat{Z}(\vec{x},t)\Big], ~\text{with}~~Q = {\rm Tr}\exp\Big[ -\int d^3x \hat{Z}(\vec{x},t)\Big],
\end{equation}
where
\begin{equation}
	\hat{Z}(\vec{x},t)=\epsilon \int_{-\infty}^t dt_1 e^{\epsilon(t_1-t)}\Big[ \beta^\nu(\vec{x},t_1)\hat{T}_{0\nu}(\vec{x},t_1)- \alpha(\vec{x},t_1)\hat{N}^0(\vec{x},t_1)-\frac{1}{2}\Omega_{\alpha\beta}(\vec{x},t_1)\hat{\Sigma}^{0\alpha\beta}(\vec{x},t_1) \Big],
\end{equation}
here, $\epsilon$ is the small infinitesimal parameter and
$\hat{T}^{\mu\nu}$, $\hat{N}^\mu$, and 
$\hat{\Sigma}^{\mu\alpha\beta}$ are operator-valued functions. 
Their statistical averages give the corresponding hydrodynamic 
fields~\eqref{Tmn},~\eqref{Nu}, and~\eqref{sigmauab}. Zubarev's 
prescription for taking the statistical average is given by
\begin{equation}
	\left< \hat{O}(\vec{x},t)\right> = \lim_{\epsilon \to 0^+} \lim_{V\to\infty}{\rm Tr}\left[\hat{\rho}(t)\hat{O}(\vec{x},t) \right].
\end{equation}
The main advantage of the NESO is that it can be decomposed into 
equilibrium and nonequilibrium parts. This can be achieved by 
integrating $\hat{Z}(\vec{x},t)$ by parts, which gives
\begin{equation}
	\hat{\rho}(t)=Q^{-1}\exp{(-\hat{A}+\hat{B})},
\end{equation}
where
\begin{align}
	\hat{A}(t)&=\int d^3x \left[ \beta^\nu(\vec{x},t)\hat{T}_{0\nu}(\vec{x},t)- \alpha(\vec{x},t)\hat{N}^0(\vec{x},t)-\frac{1}{2}\Omega_{\alpha\beta}(\vec{x},t)\hat{\Sigma}^{0\alpha\beta}(\vec{x},t) \right],\label{A_euilibrium}\\
	\hat{B}(t)&= \int d^3x \int_{-\infty}^t dt_1 ~e^{\epsilon(t_1-t)}\hat{C}(\vec{x},t_1),\label{B_noneuilibrium}\\
	\hat{C}(\vec{x},t)&=\hat{T}_{\mu\nu}(\vec{x},t)\partial^\mu\beta^\nu(\vec{x},t)-\hat{N}^\mu(\vec{x},t)\partial_\mu\alpha(\vec{x},t)-\frac{1}{2}\hat{\Sigma}^{\mu\alpha\beta}(\vec{x},t)\partial_\mu\Omega_{\alpha\beta}(\vec{x},t)\nonumber\\
	&~~~-\frac{1}{2}\Omega_{\alpha\beta}(\vec{x},t)\Big(\hat{T}^{\beta\alpha}(\vec{x},t)-\hat{T}^{\alpha\beta}(\vec{x},t)\Big),\label{c_expre}
\end{align}
here we used the conservation laws to write the nonequilibrium 
part in a covariant form. If we consider the nonequilibrium part 
$\hat{B}$ as a perturbation, the density matrix $\hat{\rho}(t)$ 
can be expanded around the local equilibrium. This expansion 
allows us to systematically account for deviations from local 
equilibrium through higher-order terms. Thus, the statistical 
average of any operator, when expanded up to the second order 
around local equilibrium, can be expressed as 
follows~\cite{HARUTYUNYAN2022168755}:
\begin{equation}
	\left< \hat{O}(x)\right> = \left< \hat{O}(x)\right>_l + \int d^4 x_1 \Big( \hat{O}(x), \hat{C}(x_1) \Big) + \int d^4 x_1 d^4 x_2 \Big( \hat{O}(x), \hat{C}(x_1), \hat{C}(x_2) \Big),
\end{equation}
here, $\int d^4 x_1 = \int d^3 x_1 \int_{-\infty}^t dt_1 
e^{\epsilon(t_1-t)}$ and the statistical averaging in local equilibrium is denoted
by the subscript $l$. The two-point and three-point 
correlation functions are defined by
\begin{align}
	\Big( \hat{O}(x), \hat{X}(x_1) \Big) &= \int_0^1 d\tau \Big< O(x) \left[ X_\tau(x_1)-\left< X_\tau(x_1) \right>_l \right] \Big>_l, \label{twopointdef}\\
	\Big( \hat{O}(x), \hat{X}(x_1), \hat{Y}(x_2) \Big) &=\frac{1}{2} \int_0^1 d\tau \int_0^1 d\lambda \Big< \tilde{T} \Big\{\hat{O}(x) \Big[ \hat{X}_\lambda (x_1) \hat{Y}_\tau(x_2)-\langle \hat{X}_\lambda (x_1)\rangle_l \hat{Y}_\tau(x_2)\nonumber\\
	&~~~-\hat{X}_\lambda (x_1) \langle \hat{Y}_\tau(x_2)\rangle_l-\langle \tilde{T} \hat{X}_\lambda (x_1) \hat{Y}_\tau(x_2)\rangle_l + 2 \langle \hat{X}_\lambda (x_1)\rangle_l \langle \hat{Y}_\tau(x_2)\rangle_l  \Big] \Big\} \Big>_l,\label{threepointdef}
\end{align}
respectively. Here, $X_\tau = e^{-\tau A} X e^{\tau A}$ and 
$\tilde{T}\{ X_\lambda Y_\tau \}$ is the anti chronological 
time-ordering operator with respect to variables $\tau$ and 
$\lambda$.

Next, we calculate the statistical average of various dissipative 
terms $(\hat{\tau}, \hat{q}, \hat{\pi}, \Pi)$ up to second order 
in gradient expansion using the NESO. We proceed by writing a tensor 
decomposition of operators $\hat{T}^{\mu\nu}$, $\hat{N}^\mu$ and 
$\hat{\Sigma}^{\mu\alpha\beta}$. The straightforward 
decomposition is given by 
\begin{align}
	\hat{T}^{\mu\nu}&=\hat{\varepsilon} u^\mu u^\nu -\hat{p}\Delta^{\mu\nu}+\hat{h}^\mu u^\nu+\hat{h}^\nu u^\mu +\hat{\pi}^{\mu\nu} +\hat{q}^\mu u^\nu-\hat{q}^\nu u^\mu + \hat{\tau}^{\mu\nu},\label{Tmno}\\
	\hat{N}^\mu &= \hat{n}u^\mu + \hat{j}^\mu, \label{Nuo}\\
	\hat{\Sigma}^{\mu\alpha\beta}&= u^\mu \hat{S}^{\alpha\beta},\label{sigmauabo}
\end{align}
where $\hat{p}$ is the actual isotropic pressure. All the 
operators, except the pressure ($\hat{p}$), can be directly 
matched with the corresponding hydrodynamic quantities by taking 
the statistical average using the full nonequilibrium statistical 
operator. That means $\varepsilon\equiv 
\big<\hat{\varepsilon}\big>$, $h^\mu\equiv 
\big<\hat{h}^\mu\big>$, $ \pi^{\mu\nu}\equiv 
\big<\hat{\pi}^{\mu\nu}\big>$, $ q^\mu\equiv 
\big<\hat{q}^\mu\big>$, $ \tau^{\mu\nu}\equiv 
\big<\hat{\tau}^{\mu\nu}\big>$, $ n\equiv \big<\hat{n}\big>$, $ 
j^{\mu}\equiv \big<\hat{j}^{\mu}\big>$, $ S^{\mu\nu}\equiv 
\big<\hat{S}^{\mu\nu}\big>$. Whereas, the statistical average $ 
\big<\hat{p}\big>\equiv p+\Pi$.
The operators can be written in terms of the projections of 
$\hat{T}^{\mu\nu}$ and $\hat{N}^\mu$,
\begin{align}
	&\hat{\varepsilon} = u^\mu u_\nu \hat{T}^{\mu\nu},~~~ \hat{p}=-\frac{1}{3}\Delta_{\mu\nu}\hat{T}^{\mu\nu},~~~ \hat{h}^{\alpha}+\hat{q}^\alpha = \Delta^\alpha_\mu u_\nu \hat{T}^{\mu\nu},~~~ \hat{h}^{\alpha}-\hat{q}^\alpha = u_\mu \Delta^\alpha_\nu \hat{T}^{\mu\nu}, \nonumber\\
	& \hat{\pi}^{\mu\nu} = \Delta_{\alpha\beta}^{\mu\nu}\hat{T}^{\alpha\beta},~~~ \hat{\tau}^{\mu\nu} = \D_{\alpha\beta}^{\mu\nu}\hat{T}^{\alpha\beta},~~~\hat{n}=u_\mu \hat{N}^\mu,~~~\hat{j}^\mu=\Delta^\mu_\nu \hat{N}^\nu,~~~  \hat{S}^{\alpha\beta}=u_\mu \hat{\Sigma}^{\mu\alpha\beta}.\label{projectionsTN}
\end{align}
The crucial quantity to evaluate for the statistical average is 
$\hat{C}$. To calculate $\hat{C}(x,t)$, we use the gradient 
decomposition of $\hat{T}^{\mu\nu}$, $\hat{N}^\mu$ and 
$\hat{\Sigma}^{\mu\alpha\beta}$ from~\eqref{Tmno},~\eqref{Nuo} 
and~\eqref{sigmauabo}, respectively. Neglecting the third-order 
gradient terms, we can rewrite~\eqref{c_expre} as
\begin{align}
	\hat{C} &= \hat{\varepsilon} D\beta -\hat{n} D\alpha -\frac{1}{2} \hat{S}^{\alpha\beta} D\Omega_{\alpha \beta}-\hat{p}\beta \theta \nonumber\\
	&~~~+ \hat{h}^\mu (\partial_\mu\beta + \beta Du_\mu) -\hat{j}^\mu \partial_\mu\alpha+ \hat{q}^\mu (\partial_\mu\beta - \beta Du_\mu+2\Omega_{\mu\nu}u^\nu)\nonumber\\ 
	&~~~+\beta \hat{\pi}^{\mu\nu}\partial_\mu u_\nu+ \hat{\tau}^{\mu\nu}(\beta\partial_\mu u_\nu+\Omega_{\mu\nu}),\label{c_first}
\end{align}
The terms $D\beta$, $D\alpha$, and $D\Omega_{\alpha \beta}$ can 
be eliminated using the equations of motion. To achieve this, we 
choose the independent variables to be $\varepsilon$, $n$, and 
$S^{\alpha\beta}$. Consequently, any thermodynamic quantity, such 
as $\beta$, can be written as a function of these variables, 
i.e., $\beta\equiv \beta(\varepsilon, n, S^{\alpha\beta})$, and 
any change in that thermodynamic quantity can be given as
\begin{equation}\label{dbeta}
	D\beta = \frac{\partial \beta}{\partial \varepsilon}\bigg|_{n,S^{\alpha\beta}} D\varepsilon + \frac{\partial \beta}{\partial n}\bigg|_{\varepsilon,S^{\alpha\beta}} Dn + \frac{\partial \beta}{\partial S^{\alpha\beta}}\bigg|_{n,\varepsilon} DS^{\alpha\beta}.
\end{equation}
The values of $D\varepsilon$, $Dn$, and $DS^{\alpha\beta}$ up to second order are 
substituted from the equations of motion~\eqref{Depsilon},~\eqref{Dn}, and~\eqref{Dsigma}, in which we fix that the terms $\Pi$, $h^\mu$, $q^\mu$, $j^\mu$, $\pi^{\mu\nu}$, and $\tau^{\mu\nu}$ only contain $\mathcal{O}(\partial^1)$ corrections. Thus, the full expression of $D\beta$, 
keeping the terms up to second order in gradient expansion, is 
given by
\begin{align}
	D\beta &=  -\theta\bigg[(\varepsilon+p)\frac{\partial \beta}{\partial \varepsilon}+n\frac{\partial \beta}{\partial n}+S^{\alpha\beta}\frac{\partial \beta}{\partial S^{\alpha\beta}}\bigg] -\frac{\partial \beta}{\partial n}\partial_\mu j^\mu -\frac{\partial \beta}{\partial S^{\alpha\beta}}\Big(2\tau^{\alpha\beta}+4q^\alpha u^\beta\Big)\nonumber\\
	&~~~-\frac{\partial \beta}{\partial \varepsilon}\Big(\Pi\theta + \partial_\mu h^\mu + u_\mu Dh^\mu + u_\nu\partial_\mu\pi^{\mu\nu}+ \partial_\mu q^\mu - u_\mu Dq^\mu  +u_\nu\partial_\mu\tau^{\mu\nu}\Big),
\end{align}
It is important to note that by excluding the third-order and 
higher-order terms, we posit that the gradient ordering of the 
term $\frac{\partial \beta}{\partial S^{\alpha\beta}}$ is 
$\mathcal{O}(\partial^1)$. This assertion can be understood by 
deriving the Maxwell relations from \eqref{tds}. The first step 
involves rewriting \eqref{tds} in terms of $\beta$, $\alpha$, and 
$\Omega_{\mu\nu}$, utilizing the connection provided by 
\eqref{connection}. Thus, we can write
\begin{equation}
	ds =\beta d\varepsilon -\alpha dn -\frac{1}{2} \Omega_{\mu\nu}dS^{\mu\nu}\label{ds}.
\end{equation}
Using~\eqref{ds}, we can find the relations
\begin{align}\label{partialrelation}
	\frac{\partial \beta}{\partial n}\bigg|_{\varepsilon,S^{\alpha\beta}} = -\frac{\partial \alpha}{\partial \varepsilon}\bigg|_{n,S^{\alpha\beta}},~~ \frac{\partial \beta}{\partial S^{\alpha\beta}}\bigg|_{\varepsilon,n} = -\frac{1}{2}\frac{\partial \Omega_{\alpha\beta}}{\partial \varepsilon}\bigg|_{n,S^{\alpha\beta}},~~
	\frac{\partial \alpha}{\partial S^{\alpha\beta}}\bigg|_{\varepsilon,n} = \frac{1}{2}\frac{\partial \Omega_{\alpha\beta}}{\partial n}\bigg|_{\varepsilon,S^{\alpha\beta}}.
\end{align}
Given that $\Omega_{\alpha\beta}\sim \mathcal{O}(\partial^1)$,  
it follows that the partial derivative $\frac{\partial 
\Omega_{\alpha\beta}}{\partial \varepsilon}$ is also of the order 
$\mathcal{O}(\partial^1)$. 

Similarly, we can also calculate $D\alpha$ and $D\Omega_{\alpha \beta}$ given by
\begin{align}
	D\alpha &=  -\theta\bigg[(\varepsilon+p)\frac{\partial \alpha}{\partial \varepsilon}+n\frac{\partial \alpha}{\partial n}+S^{\alpha\beta}\frac{\partial \alpha}{\partial S^{\alpha\beta}}\bigg] -\frac{\partial \alpha}{\partial n}\partial_\mu j^\mu -\frac{\partial \alpha}{\partial S^{\alpha\beta}}\Big(2\tau^{\alpha\beta}+4q^\alpha u^\beta\Big) \nonumber\\
	&~~~ -\frac{\partial \alpha}{\partial \varepsilon}\Big(\Pi\theta + \partial_\mu h^\mu + u_\mu Dh^\mu + u_\nu\partial_\mu\pi^{\mu\nu}+ \partial_\mu q^\mu - u_\mu Dq^\mu  +u_\nu\partial_\mu\tau^{\mu\nu}\Big),
\end{align}
and,
\begin{align}	
	D\Omega_{\alpha\beta} &= -\theta\bigg[(\varepsilon+p)\frac{\partial \Omega_{\alpha\beta}}{\partial \varepsilon}+n\frac{\partial \Omega_{\alpha\beta}}{\partial n}+S^{\mu\nu}\frac{\partial \Omega_{\alpha\beta}}{\partial S^{\mu\nu}}\bigg] -\frac{\partial \Omega_{\alpha\beta}}{\partial n}\partial_\mu j^\mu -\frac{\partial \Omega_{\alpha\beta}}{\partial S^{\mu\nu}}\Big(2\tau^{\mu\nu}+4q^\mu u^\nu\Big) \nonumber\\
	&~~~ -\frac{\partial \Omega_{\alpha\beta}}{\partial \varepsilon}\Big(\Pi\theta + \partial_\mu h^\mu + u_\mu Dh^\mu + u_\nu\partial_\mu\pi^{\mu\nu}+ \partial_\mu q^\mu - u_\mu Dq^\mu  +u_\nu\partial_\mu\tau^{\mu\nu}\Big).
\end{align}
Using the values of $D\beta$, $D\alpha$ and $D\Omega_{\alpha 
\beta}$ together with the relations~\eqref{partialrelation}, and extracting the 
value of $Du_\alpha$ from~\eqref{Du}, we can express 
the nonequilibrium part $(\hat{C})$ of the nonequilibrium 
statistical operator in the Landau frame. Thus, we obtain
\begin{align}
	\hat{C} &= -\hat{P}^* \beta \theta  -\hat{j}^\mu \partial_\mu\alpha +\hat{q}^\mu (\partial_\mu\beta - \beta h^{-1}\nabla_\mu p + 2\Omega_{\mu\nu} u^\nu)+\beta \hat{\pi}^{\mu\nu}\partial_\mu u_\nu +\hat{\tau}^{\mu\nu}(\beta \partial_\mu u_\nu+\Omega_{\mu\nu})\nonumber\\
	&~~~ +\hat{q}^\alpha \beta h^{-1}\Big[ \Pi Du_\alpha -\nabla_\alpha\Pi  + \Delta_{\alpha\nu}\partial_\mu\pi^{\mu\nu} -q_\alpha\theta+ q^\mu\partial_\mu u_\alpha - \Delta_{\alpha\nu} Dq^\nu + \Delta_{\alpha\nu}\partial_\mu\tau^{\mu\nu} \Big]\nonumber\\
	&~~~ -\hat{\beta}^* \Big(\Pi\theta + u_\nu\partial_\mu\pi^{\mu\nu}+ \partial_\mu q^\mu - u_\mu Dq^\mu  +u_\nu\partial_\mu\tau^{\mu\nu}\Big)+ \hat{\alpha}^* \partial_\mu j^\mu + \hat{\Omega}_{\mu\nu}^{*}(\tau^{\mu\nu}+2q^\mu u^\nu).\label{finalC}
\end{align}
Here, we set $h_\mu=0$ for the Landau frame, and we have defined
\begin{align}\label{totalpressure}
	\hat{P}^* = (\hat{p}-\hat{\varepsilon}\Gamma - \hat{n}\delta) - \hat{S}^{\alpha\beta} \mathcal{K}_{\alpha\beta}= \hat{p}^* - \hat{S}^{\alpha\beta} \mathcal{K}_{\alpha\beta} ,
\end{align}
and the new parameters as
\begin{align}
	  \hat{\beta}^* &= \hat{\varepsilon}\frac{\partial \beta}{\partial \varepsilon}+ \hat{n} \frac{\partial \beta}{\partial n}+	\hat{S}^{\alpha\beta}\frac{\partial \beta}{\partial S^{\alpha\beta}}, \nonumber\\
	   \hat{\alpha}^*&=\hat{\varepsilon}\frac{\partial \alpha}{\partial \varepsilon}+\hat{n}\frac{\partial \alpha}{\partial n}+	\hat{S}^{\alpha\beta}\frac{\partial \alpha}{\partial S^{\alpha\beta}} ,\nonumber\\
	   \hat{\Omega}_{\mu\nu}^{*}&=\hat{\varepsilon} \frac{\partial \Omega_{\mu\nu}}{\partial \varepsilon}+ \hat{n} \frac{\partial \Omega_{\mu\nu}}{\partial n}+\hat{S}^{\alpha\beta} \frac{\partial \Omega_{\alpha\beta}}{\partial S^{\mu\nu}}.
\end{align}
We have also defined,
\begin{equation}
	\Gamma=\frac{\partial p}{\partial \varepsilon}\bigg|_{n, S^{\mu\nu}}, ~~ \delta=\frac{\partial p}{\partial n}\bigg|_{\varepsilon, S^{\mu\nu}} ~~\text{and,}~~ \mathcal{K}_{\alpha\beta}=\frac{\partial p}{\partial S^{\alpha\beta}}\bigg|_{\varepsilon,n}	.
\end{equation}
Next, we separate $\hat{C}$ into two parts, 
$\hat{C}=\hat{C}_F+\hat{C}_S$, containing first-order and 
second-order gradient terms, respectively. $\hat{C}_F$ and $\hat{C}_S$ are 
\begin{align}
	\hat{C}_F &= -\hat{p}^*\beta\theta -\hat{j}^\mu\nabla_\mu\alpha + \beta\hat{q}^\mu M_\mu +\beta \hat{\pi}^{\mu\nu}\sigma_{\mu\nu} +\beta\hat{\tau}^{\mu\nu}\xi_{\mu\nu}, \label{CF}\\
	\hat{C}_S &=  \beta\sum_{i=1,2}\Big[\big(\hat{\mathfrak{D}}_i\partial^i_{\varepsilon n}\beta\big)\mathcal{X} +\big(\hat{\mathfrak{D}}_i\partial^i_{\varepsilon n}\alpha\big)\mathcal{Y} + \big(\hat{\mathfrak{D}}_i\partial^i_{\varepsilon n}\Omega_{\mu\nu}\big) \mathcal{Z}^{\mu\nu}\Big]+ \beta \hat{q}^\alpha\mathcal{Q}_\alpha + \hat{S}^{\mu\nu}\mathcal{K}_{\mu\nu}\beta \theta. \label{CS}
\end{align} 
Here, $\hat{\mathfrak{D}}_i\equiv(\hat{\varepsilon},\hat{n})$ and 
$\partial^i_{\varepsilon n}\equiv 
\big(\frac{\partial}{\partial\varepsilon}, 
\frac{\partial}{\partial n}\big)$ and we defined 
\begin{align}
	\mathcal{Q}_\alpha &= h^{-1}\Big[ \Pi Du_\alpha -\nabla_\alpha\Pi  + \Delta_{\alpha\nu}\partial_\mu\pi^{\mu\nu} -q_\alpha\theta+ q^\mu\partial_\mu u_\alpha - \Delta_{\alpha\nu} Dq^\nu + \Delta_{\alpha\nu}\partial_\mu\tau^{\mu\nu} \Big], \\
	\mathcal{X} &=-\beta^{-1}\Big(\Pi\theta + u_\nu\partial_\mu\pi^{\mu\nu}+ \partial_\mu q^\mu - u_\mu Dq^\mu  +u_\nu\partial_\mu\tau^{\mu\nu}\Big), \\
	\mathcal{Y} &= \beta^{-1}\partial_\mu j^\mu, \\
	\mathcal{Z}^{\mu\nu}&= \beta^{-1}\Big( \tau^{\mu\nu}+2q^\mu u^\nu \Big).
\end{align}
Employing~\eqref{CF} and~\eqref{CS}, we can rewrite the 
statistical average of any operator in terms of gradient 
expansion as
\begin{equation}\label{gradient_expansion}
	\left< \hat{O}(x)\right> = \left< \hat{O}(x)\right>_l + \left< \hat{O}(x)\right>_1 + \left< \hat{O}(x)\right>_2,
\end{equation}
where $\left< \hat{O}(x)\right>_1$ and $\left< 
\hat{O}(x)\right>_2$ are the first-order and second-order 
gradient terms, respectively. The first-order contribution is 
calculated assuming that the thermodynamic forces remain 
constant within the correlation length. This yields the local 
contribution from the two-point correlation function. The 
resulting expression is given by 
\begin{equation}
	\left< \hat{O}(x)\right>_1 = \int d^4 x_1 \Big( \hat{O}(x), \hat{C}_F(x_1) \Big)\Big|_{local}.\label{FOC}
\end{equation}
Whereas, the second-order contribution to the statistical average 
can originate from three possible ways, given by
\begin{equation}
	\left< \hat{O}(x)\right>_2 = \left< \hat{O}(x)\right>_2^{(2),NL} + \left< \hat{O}(x)\right>_2^{(2),ET} +\left< \hat{O}(x)\right>_2^{(3)} ,
\end{equation}
where,
\begin{align}
	\left< \hat{O}(x)\right>_2^{(2),NL} &=  \int d^4 x_1 \Big( \hat{O}(x), \hat{C}_F(x_1) \Big)-\left< \hat{O}(x)\right>_1, \label{nlc}\\
	\left< \hat{O}(x)\right>_2^{(2),ET} &= \int d^4 x_1 \Big( \hat{O}(x), \hat{C}_S(x_1) \Big)\Big|_{local}, \label{etc}\\
	\left< \hat{O}(x)\right>_2^{(3)}  &= \int d^4 x_1 d^4 x_2 \Big( \hat{O}(x), \hat{C}_F(x_1), \hat{C}_F(x_2) \Big)\Big|_{local} , \label{tpc}
\end{align}
here, $\left< \hat{O}(x)\right>_2^{(2),NL}$ is the nonlocal ($NL$)
correction from the two-point correlation function, $\left< 
\hat{O}(x)\right>_2^{(2),ET}$ is the local correction from 
the two-point correlation function using external thermodynamic 
forces ($ET$), and $\left< \hat{O}(x)\right>_2^{(3)}$ is the local 
correction from the three-point correlation function. The meaning 
of local and nonlocal are explained in the next section.


\section{First-order spin hydrodynamics}\label{FO_SH}
The 
transport coefficients emerging from first-order spin 
hydrodynamics have been calculated using the density operator 
formalism \cite{Hu:2021lnx}.  In this section, we provide a 
concise overview of the calculations pertinent to first-order 
spin hydrodynamics. We begin by calculating the average value for 
the rotational stress tensor $\tau_{\mu\nu}$, the antisymmetric 
counterpart of the $\pi_{\mu\nu}$. Using~\eqref{FOC}, the 
first-order gradient correction to $\big< \hat{\tau}_{\mu\nu}(x) 
\big>$ is given by
\begin{equation}
	\big< \hat{\tau}_{\mu\nu}(x) \big>_1 = \int d^4 x_1 \Big( \hat{\tau}_{\mu\nu}(x), \hat{C}_F(x_1) \Big)\Big|_{local}.
\end{equation}
Now, we take the help of Curie's principle, that the correlation 
between the operators of different ranks vanishes in an isotropic 
media. Hence, the only non zero correlation for the $\big< 
\hat{\tau}_{\mu\nu}(x) \big>_1$ is 
\begin{equation}\label{taufo}
	\big< \hat{\tau}_{\mu\nu}(x) \big>_1 = \int d^4 x_1 \Big( \hat{\tau}_{\mu\nu}(x), \hat{\tau}_{\rho\sigma}(x_1) \Big)\beta(x_1)\xi^{\rho\sigma}(x_1)\Big|_{local} + \int d^4 x_1 \Big( \hat{\tau}_{\mu\nu}(x), \hat{\pi}_{\rho\sigma}(x_1) \Big)\beta(x_1)\sigma^{\rho\sigma}(x_1)\Big|_{local}.
\end{equation}
Also, in an isotropic medium, we can write (see 
Appendix~\ref{correlations_projectors} Eq.~\eqref{four_ten_deco2})
\begin{align}
	\Big( \hat{\tau}_{\mu\nu}(x), \hat{\tau}_{\rho\sigma}(x_1) \Big)&=\frac{1}{3}\D_{\mu\nu\rho\sigma}\Big( \hat{\tau}^{\lambda\eta}(x), \hat{\tau}_{\lambda\eta}(x_1) \Big),\label{asydec}\\
	\Big( \hat{\tau}_{\mu\nu}(x), \hat{\pi}_{\rho\sigma}(x_1) \Big) &= 0.\label{asydec2}
\end{align}
Further, we use the Taylor expansion to expand 
$\beta(x_1)\xi^{\rho\sigma}(x_1)$ around $x_1=x$. We have
\begin{equation}\label{tayextau}
	\beta(x_1)\xi^{\rho\sigma}(x_1) = \beta(x)\xi^{\rho\sigma}(x)+(x_1-x)^\alpha \frac{\partial}{\partial x_1^\alpha}\left[\beta(x_1)\xi^{\rho\sigma}(x_1)\right]\bigg|_x +\cdots .
\end{equation}
Here, we like to emphasize the gradient ordering in the second term. In this context, $(x_1-x)^\alpha$ represents the 
correlation length, which is comparable to the microscopic 
length, $i.e.$, the mean free path of the particles 
$(\lambda)$. Consequently, the term $(x_1-x)^\alpha 
\frac{\partial}{\partial x_1^\alpha}\sim \frac{\lambda}{L}$, 
where $L$ is the macroscopic scale. The ratio $\frac{\lambda}{L}$ 
known as Knudsen number. Therefore, the second term 
in~\eqref{tayextau} is of the second order in hydrodynamic 
gradient expansion. This term is neglected in first-order 
hydrodynamics calculation. The first term in~\eqref{tayextau}, 
evaluated at $x_1=x$ is the local term. All the higher-order terms are called nonlocal terms. 
Substituting~\eqref{asydec},~\eqref{asydec2}, 
and~\eqref{tayextau} in~\eqref{taufo}, and keeping only the 
first-order gradient terms, we obtain
\begin{equation}
	\tau_{\mu\nu}(x) \equiv \big< \hat{\tau}_{\mu\nu}(x) \big>_1 =  \frac{1}{3} \beta(x)\xi_{\mu\nu}(x)\int d^4 x_1\Big( \hat{\tau}^{\lambda\eta}(x), \hat{\tau}_{\lambda\eta}(x_1) \Big).\label{taufirst}
\end{equation}
By employing Curie's principle, a similar approach is applied to 
the remaining dissipative quantities. Thus, we can write
\begin{align}
	q_\mu(x)\equiv \big< \hat{q}_\mu(x) \big>_1 &= \frac{1}{3}\beta(x)M_\mu(x)\int d^4 x_1 \Big( \hat{q}^\lambda(x), \hat{q}_\lambda(x_1) \Big),\label{qfirst}\\
	\pi_{\mu\nu}(x) \equiv \big< \hat{\pi}_{\mu\nu}(x) \big>_1 &= \frac{1}{5} \beta(x)\sigma_{\mu\nu}(x)\int d^4 x_1\Big( \hat{\pi}^{\lambda\eta}(x), \hat{\pi}_{\lambda\eta}(x_1) \Big).\label{pifirst}
\end{align}
However, the bulk viscous pressure $(\Pi)$ is defined by
\begin{equation}
	\Pi = \big< \hat{p} \big>-p(\varepsilon,n,S^{\alpha\beta}),
\end{equation} 
and up to the first order in gradient expansion, we obtain~(see 
Appendix~\ref{bulk} Eq.~\eqref{bulk_exp_complete})
\begin{equation}
	\Pi=\big< \hat{p}^*(x) \big>_1,
\end{equation}
which is given by
\begin{align}	
	\Pi=\big< \hat{p}^*(x) \big>_1 &= -\beta(x)\theta(x)\int d^4 x_1 \Big( \hat{p}^*(x), \hat{p}^*(x_1) \Big),\label{pfirst}
\end{align}
here, we followed a procedure similar to the one outlined while 
calculating~\eqref{taufirst}. Comparing the 
equations~\eqref{taufirst},~\eqref{qfirst},~\eqref{pifirst} 
and~\eqref{pfirst} with~\eqref{transport}, we can deduce the 
first-order transport coefficients, given by
\begin{align}
	\gamma &= \frac{\beta}{6} \int d^4 x_1 \Big( \hat{\tau}^{\lambda\eta}(x), \hat{\tau}_{\lambda\eta}(x_1) \Big),\label{gamma_first}\\
	\lambda &= -\frac{\beta}{3} \int d^4 x_1 \Big( \hat{q}^\lambda(x), \hat{q}_\lambda(x_1) \Big),\label{lambda_first}\\
	\eta &= \frac{\beta}{10} \int d^4 x_1 \Big( \hat{\pi}^{\lambda\eta}(x), \hat{\pi}_{\lambda\eta}(x_1) \Big),\label{eta_first}\\
	\zeta &= \beta \int d^4 x_1 \Big( \hat{p}^*(x), \hat{p}^*(x_1) \Big).\label{zeta_first}
\end{align}

\section{Second-order spin hydrodynamics}
Relativistic hydrodynamics, which considers only the symmetric 
canonical energy-momentum tensor and excludes the spin degree of freedom, has been extensively studied up to the second-order using the density operator 
formalism~\cite{HARUTYUNYAN2022168755}. We extend this formalism 
to include the antisymmetric part of the energy-momentum tensor and 
derive the second-order spin hydrodynamics. We start by 
calculating the second-order corrections to the rotational stress 
tensor.

\subsection{Rotational Stress Tensor ($\tau_{\mu\nu}$)}

The second-order correction to $\big< \hat{\tau}_{\mu\nu}(x) 
\big>$ can be decomposed into three parts (see~\eqref{FOC}), given by
\begin{equation}
	\big< \hat{\tau}_{\mu\nu}(x) \big>_2 = \big< \hat{\tau}_{\mu\nu}(x) \big>_2^{(2),NL} + \big< \hat{\tau}_{\mu\nu}(x) \big>_2^{(2),ET} + \big< \hat{\tau}_{\mu\nu}(x) \big>_2^{(3)}.
\end{equation}
The general definition of these three terms is defined 
in~\eqref{nlc},~\eqref{etc}, and~\eqref{tpc}. Since we are 
interested in the correlation in isotropic media, these terms 
can be further simplified using Curie's principle. Thus, 
\begin{align}
	\big< \hat{\tau}_{\mu\nu}(x) \big>_2^{(2),NL}&=  \int d^4 x_1 \Big( \hat{\tau}_{\mu\nu}(x), \hat{\tau}_{\alpha\beta}(x_1) \Big)\beta(x_1)\xi^{\alpha\beta}(x_1)-\big< \hat{\tau}_{\mu\nu}(x) \big>_1,\label{tau_nl}\\
	\big< \hat{\tau}_{\mu\nu}(x) \big>_2^{(2),ET}&=  \int d^4 x_1 \Big( \hat{\tau}_{\mu\nu}(x), \hat{S}_{\alpha\beta}(x_1) \Big)\mathcal{K}^{\alpha\beta}(x_1)\beta(x_1) \theta (x_1)\Big|_{local},\label{tau_et}\\
	\big< \hat{\tau}_{\mu\nu}(x) \big>_2^{(3)}	&= \int d^4 x_1 d^4 x_2 \Big( \hat{\tau}_{\mu\nu}(x), \hat{C}_F(x_1),\hat{C}_F(x_2) \Big)\Big|_{local},\label{tau_3c}
\end{align}
where, $\hat{C}_F$ is given by the equation~\eqref{CF}.

\subsubsection{Nonlocal correction from two-point correlation function: $\big< \hat{\tau}_{\mu\nu}(x) \big>_2^{(2),NL}$}\label{tau_nlc}
To calculate the nonlocal correction from the two-point 
correlation function, we first generalize Eq~\eqref{asydec}, 
which can be given by
\begin{equation}
	\Big( \hat{\tau}_{\mu\nu}(x), \hat{\tau}_{\rho\sigma}(x_1) \Big)=\frac{1}{3}\D_{\mu\nu\rho\sigma}(x,x_1)\Big( \hat{\tau}^{\lambda\eta}(x), \hat{\tau}_{\lambda\eta}(x_1) \Big).\label{tauprodec}
\end{equation}
This is the nonlocal generalization of two-point correlators, 
with the nonlocal projector defined as (see 
\ref{non-local_projectors})
\begin{equation}
	\D_{\mu\nu\rho\sigma}(x,x_1)=\D_{\mu\nu\alpha\beta}(x)\D^{\alpha\beta}_{\rho\sigma}(x_1).\label{nlaspro}
\end{equation}
In order to extract nonlocal terms from~\eqref{tau_nl}, we have 
to express $\hat{\tau}^{\lambda\sigma}$ and 
$\hat{\tau}_{\lambda\sigma}$ in terms of energy-momentum tensor, 
and then we expand all the hydrodynamic quantities around 
$x_1=x$. From~\eqref{projectionsTN}, we can write
\begin{align}
	\hat{\tau}^{\lambda\eta} (x) &= \D^{\lambda\eta}_{\gamma\delta}(x) \hat{T}^{\gamma\delta}(x),\label{tauproj1}\\
	\hat{\tau}_{\lambda\eta} (x_1) &= \D_{\lambda\eta \alpha\beta}(x_1) \hat{T}^{\alpha\beta}(x_1).\label{tauproj2}
\end{align}
Substituting~\eqref{tauproj1} and~\eqref{tauproj2} together 
with~\eqref{tauprodec} and~\eqref{nlaspro} in~\eqref{tau_nl}, we 
obtain
\begin{align}\label{tau_2_a}
	\big< \hat{\tau}_{\mu\nu}(x) \big>_2^{(2),NL} &=\frac{1}{3}\D_{\mu\nu\rho\sigma}(x)\int d^4 x_1 \Big( \hat{T}^{\gamma\delta}(x), \hat{T}^{\alpha\beta}(x_1) \Big)\D_{\gamma \delta \alpha\beta}(x,x_1)\beta(x_1)\xi^{\rho\sigma}(x_1)-\big< \hat{\tau}_{\mu\nu}(x) \big>_1.
\end{align}
Furthermore, the Taylor's expansion of $\D_{\gamma \delta 
	\alpha\beta}(x,x_1)\beta(x_1)\xi^{\rho\sigma}(x_1)$ around 
$x_1=x$ is given by 
\begin{align}\label{asy_expa}
	\D_{\gamma \delta \alpha\beta}(x,x_1)\beta(x_1)\xi^{\rho\sigma}(x_1) &= \D_{\gamma \delta \alpha\beta}(x)\beta(x)\xi^{\rho\sigma}(x) + (x_1-x)^\lambda \Bigg[ \D_{\gamma \delta \alpha\beta}(x)\frac{\partial}{\partial x_1^\lambda}\big\{\beta(x_1)\xi^{\rho\sigma}(x_1) \big\}\Big|_{x_1=x}\nonumber\\
	&~~~ +\beta(x)\xi^{\rho\sigma}(x)\frac{\partial}{\partial x_1^\lambda}\big\{\D_{\gamma \delta \alpha\beta}(x,x_1) \big\}\Big|_{x_1=x} \Bigg],
\end{align}
where the first-order derivative of $\D_{\gamma \delta 
\alpha\beta}(x,x_1)$ is given by {(see 
Eq.~\eqref{asy_derivative} in~\ref{non-local_projectors})}
\begin{equation}\label{deri_asy}
	\frac{\partial}{\partial x_1^\lambda}\D_{\mu\nu\rho\sigma}(x,x_1)\bigg|_{x_1=x} = -\Big[ \D_{\mu\nu\rho\tau}(x)u_\sigma(x)+\D_{\mu\nu\tau\sigma}(x)u_\rho(x) \Big]\frac{\partial u^\tau(x_1)}{\partial x_1^\lambda}\bigg|_{x_1=x}.
\end{equation}
Substituting~\eqref{asy_expa} in~\eqref{tau_2_a}, we find that 
the leading-order term in~\eqref{asy_expa} cancels $\big< 
\hat{\tau}_{\mu\nu}(x) \big>_1$ in~\eqref{tau_2_a}, with the help 
of~\eqref{taufirst}. Thus, we obtain
\begin{align}\label{tau_2_a_2}
	\big< \hat{\tau}_{\mu\nu}(x) \big>_2^{(2),NL} &= \bigg[ \frac{1}{3} \D_{\mu\nu\rho\sigma}(x)\D_{\gamma \delta \alpha\beta}(x) \frac{\partial}{\partial x_1^\lambda}\Big\{ \beta(x_1)\xi^{\rho\sigma}(x_1) \Big\}_{x_1=x} \int d^4 x_1  \Big( \hat{T}^{\gamma\delta}(x), \hat{T}^{\alpha\beta}(x_1) \Big) (x_1-x)^\lambda \bigg]\nonumber\\
	&~~~ -\bigg[ \frac{1}{3} \beta(x)\xi_{\mu\nu}(x)\frac{\partial u^\tau(x_1)}{\partial x_1^\lambda}\bigg|_{x_1=x} \Big\{ \D_{\gamma \delta \alpha\tau}(x)u_\beta(x)+\D_{\gamma \delta \tau\beta}(x)u_\alpha(x) \Big\}\nonumber\\
	&~~~ \times \int d^4 x_1  \Big( \hat{T}^{\gamma\delta}(x), \hat{T}^{\alpha\beta}(x_1) \Big) (x_1-x)^\lambda \bigg].
\end{align}
Next, we approximate $u_\beta(x)\simeq u_\beta(x_1)$ using the 
leading-order approximation, because the terms 
in~\eqref{tau_2_a_2} are already multiplied with the second-order 
terms. Thus, using~\eqref{projectionsTN} in the Landau frame, we 
find that the second part of the equation~\eqref{tau_2_a_2} gives 
the two types of correlation,
\begin{equation}\label{zerocorrtau}
	\int d^4 x_1 \Big( \hat{\tau}_{\alpha\tau}(x), \hat{q}^{\alpha}(x_1) \Big), ~~~\text{and}~~~\int d^4 x_1 \Big( \hat{\tau}_{\alpha\tau}(x), \hat{\varepsilon}(x_1)u^{\alpha}(x_1) \Big), 
\end{equation}
which can be safely set to zero, by revoking Curie's principle 
and using the property $u^\alpha\hat{\tau}_{\alpha\tau}=0$. Also, 
in the first-order gradient approximation, we can write,
\begin{equation}
	\D_{\gamma \delta \alpha\beta} (x)\int d^4 x_1 \Big( \hat{T}^{\gamma\delta}(x), \hat{T}^{\alpha\beta}(x_1) \Big) = \int d^4 x_1  \Big( \hat{\tau}^{\lambda\sigma}(x), \hat{\tau}_{\lambda\sigma}(x_1) \Big),\label{tau_sim1}
\end{equation}
Substituting~\eqref{tau_sim1} in~\eqref{tau_2_a_2} and 
using~\eqref{zerocorrtau}, we get 
\begin{equation}\label{tau_2_a_3}
	\big< \hat{\tau}_{\mu\nu}(x) \big>_2^{(2),NL} = 2[\beta(x)]^{-1} \D_{\mu\nu\rho\sigma}(x) L^\lambda(x)  \frac{\partial}{\partial x_1^\lambda}\Big\{ \beta(x_1)\xi^{\rho\sigma}(x_1) \Big\}\bigg|_{x_1=x}.
\end{equation}
Here, we defined,
\begin{equation}\label{def_trans_a}
	L^\lambda(x) = \frac{\beta(x)}{6} \int d^4 x_1  \Big( \hat{\tau}^{\alpha\beta}(x), \hat{\tau}_{\alpha\beta}(x_1) \Big) (x_1-x)^\lambda,
\end{equation}
which can be further simplified in the following form (see 
Eq.~\eqref{freq_dep_trans} in Appendix~\ref{correlators})
\begin{equation}\label{def_trans_a2}
	L^\lambda = \gamma_\omega u^\lambda,
\end{equation}
where, 
\begin{equation}
	\gamma_\omega = i\lim_{\omega\to0}\frac{d}{d\omega}\gamma_{\hat{\tau}\hat{\tau}}(\omega).
\end{equation}
Here $\gamma_{\hat{\tau}\hat{\tau}}(\omega)$ is the frequency-dependent transport coefficient, given by (see 
Appendix~\ref{correlators})
\begin{equation}
	\gamma_{\hat{\tau}\hat{\tau}}(\omega)=\frac{\beta}{6} \int d^4x_1 e^{i\omega(t-t_1)}\Big( \hat{\tau}^{\alpha\beta}(x), \hat{\tau}_{\alpha\beta}(x_1) \Big).
\end{equation}
Combining~\eqref{tau_2_a_3},~\eqref{def_trans_a} 
and~\eqref{def_trans_a2}, and using the first-order approximation 
$D\beta\simeq\Gamma\beta\theta$, we obtain
\begin{equation}\label{finaltau2nl}
	\big< \hat{\tau}_{\mu\nu}\big>_2^{(2),NL} = 2\gamma_\omega \left[ \D_{\mu\nu\rho\sigma} D\xi^{\rho\sigma} +\theta \Gamma \xi_{\mu\nu} \right],
\end{equation}
where we removed the argument $x$ from all quantities.

\subsubsection{Local correction from two-point correlation function using extended thermodynamic forces: $\big< \hat{\tau}_{\mu\nu}(x) \big>_2^{(2),ET}$}

Since $\hat{C}_S$ contains an antisymmetric tensor operator 
$\hat{S}_{\alpha\beta}$, the term~\eqref{tau_et} is nonzero for 
a rotational stress tensor. To solve this term, we can expand 
$\mathcal{K}^{\alpha\beta}(x_1)\beta(x_1) \theta (x_1) $ around 
$x_1=x$, in a manner similar to how we solved for $\big< 
\hat{\tau}_{\mu\nu}(x) \big>_1$. From~\eqref{tau_et}, we have
\begin{align}\label{finaltau2et}
	\big< \hat{\tau}_{\mu\nu}(x) \big>_2^{(2),ET}	
	&=  \int d^4 x_1 \Big( \hat{\tau}_{\mu\nu}(x), \hat{S}_{\alpha\beta}(x_1) \Big)\mathcal{K}^{\alpha\beta}(x_1)\beta(x_1) \theta (x_1)\bigg|_{local},\nonumber\\
	&= \gamma_{{}_{\tau S}}(x) \mathcal{K}_{\mu\nu}(x) \theta (x) ,
\end{align}
where we defined
\begin{equation}
	\gamma_{{}_{\tau S}}(x) = \frac{\beta(x)}{3}\int d^4 x_1 \Big( \hat{\tau}_{\lambda\sigma}(x), \hat{S}^{\lambda\sigma}(x_1) \Big).
\end{equation}
Also, we have used (see Appendix~\ref{correlations_projectors}), 
\begin{equation}
	\Big( \hat{\tau}_{\mu\nu}(x), \hat{S}_{\alpha\beta}(x_1) \Big) = \frac{1}{3} \D_{\mu\nu\alpha\beta}\Big( \hat{\tau}_{\lambda\sigma}(x), \hat{S}^{\lambda\sigma}(x_1) \Big).
\end{equation}


\subsubsection{Local correction from three-point correlation function: $\big< \hat{\tau}_{\mu\nu}(x) \big>_2^{(3)}$}\label{tau3p}
Using~\eqref{tpc}, this correction term is given by
\begin{equation}\label{tau_2_c}
	\big< \hat{\tau}_{\mu\nu}(x) \big>_2^{(3)}	= \int d^4 x_1 d^4 x_2 \Big( \hat{\tau}_{\mu\nu}(x), \hat{C}_F(x_1),\hat{C}_F(x_2) \Big) \Big|_{local},
\end{equation}
where, 
\begin{equation}
	\hat{C}_F(x) = -\hat{p}^*\beta\theta -\hat{j}_\mu \nabla^\mu \alpha + \beta\hat{q}_\mu M^\mu+\beta \hat{\pi}_{\mu\nu}\sigma^{\mu\nu} +\beta\hat{\tau}_{\mu\nu}\xi^{\mu\nu},
\end{equation}
and $M^\mu=\beta^{-1}\partial^\mu\beta - h^{-1}\nabla^\mu p + 
2\beta^{-1}\Omega^{\mu\nu} u_\nu$. 

In order to calculate the local contribution 
from~\eqref{tau_2_c}, we expand the thermodynamic forces around 
$x_1=x$ and $x_2=x$, and collect only the leading-order terms, which 
is equivalent to replacing the argument of thermodynamic forces 
with $x$. Thus, substituting the value of $\hat{C}_F$ 
in~\eqref{tau_2_c}, we get ten types of nonzero correlations, 
given by
\begin{align}
	\mathcal{A}_{1\mu\nu}(x) &= \beta^2(x)\xi^{\rho\sigma}(x)\xi^{\alpha\beta}(x)\int d^4 x_1 d^4 x_2 \Big( \hat{\tau}_{\mu\nu}(x), \hat{\tau}_{\rho\sigma}(x_1),\hat{\tau}_{\alpha\beta}(x_2) \Big),  \label{termtauc1} \\
	\mathcal{A}_{2\mu\nu}(x) &= \beta^2(x)\sigma^{\rho\sigma}(x)\sigma^{\alpha\beta}(x)\int d^4 x_1 d^4 x_2 \Big( \hat{\tau}_{\mu\nu}(x), \hat{\pi}_{\rho\sigma}(x_1),\hat{\pi}_{\alpha\beta}(x_2) \Big), \label{termtauc2}\\
	\mathcal{A}_{3\mu\nu}(x) &= \beta^2(x)\sigma^{\rho\sigma}(x)\xi^{\alpha\beta}(x)\int d^4 x_1 d^4 x_2 \Big( \hat{\tau}_{\mu\nu}(x), \hat{\pi}_{\rho\sigma}(x_1),\hat{\tau}_{\alpha\beta}(x_2) \Big),\label{termtauc3}\\
	\mathcal{A}_{4\mu\nu}(x) &= \beta^2(x)\xi^{\rho\sigma}(x)\sigma^{\alpha\beta}(x)\int d^4 x_1 d^4 x_2 \Big( \hat{\tau}_{\mu\nu}(x), \hat{\tau}_{\rho\sigma}(x_1),\hat{\pi}_{\alpha\beta}(x_2) \Big),\label{termtauc4}\\
	\mathcal{A}_{5\mu\nu}(x) &= -\beta^2(x)\xi^{\rho\sigma}(x)\theta(x)\int d^4 x_1 d^4 x_2 \Big( \hat{\tau}_{\mu\nu}(x), \hat{\tau}_{\rho\sigma}(x_1),\hat{p}^*(x_2) \Big), \label{termtauc5}\\
	\mathcal{A}_{6\mu\nu}(x) &= -\beta^2(x)\xi^{\alpha\beta}(x)\theta(x)\int d^4 x_1 d^4 x_2 \Big( \hat{\tau}_{\mu\nu}(x), \hat{p}^*(x_1), \hat{\tau}_{\alpha\beta}(x_2) \Big),\label{termtauc6}\\
	\mathcal{A}_{7\mu\nu}(x) &= \beta^2(x)M^\alpha(x)M^\beta(x)\int d^4 x_1 d^4 x_2 \Big( \hat{\tau}_{\mu\nu}(x), \hat{q}_\alpha(x_1), \hat{q}_\beta(x_2) \Big),\label{termtauc7}\\
	\mathcal{A}_{8\mu\nu}(x) &= \nabla^\alpha\alpha(x)\nabla^\beta\alpha(x)\int d^4 x_1 d^4 x_2 \Big( \hat{\tau}_{\mu\nu}(x), \hat{j}_\alpha(x_1), \hat{j}_\beta(x_2) \Big),\label{termtauc8}\\
	\mathcal{A}_{9\mu\nu}(x) &= -\beta(x)\nabla^\alpha\alpha(x)M^\beta(x)\int d^4 x_1 d^4 x_2 \Big( \hat{\tau}_{\mu\nu}(x), \hat{j}_\alpha(x_1), \hat{q}_\beta(x_2) \Big),\label{termtauc9}\\
	\mathcal{A}_{10\mu\nu}(x) &= -\beta(x)M^\alpha(x)\nabla^\beta\alpha(x)\int d^4 x_1 d^4 x_2 \Big( \hat{\tau}_{\mu\nu}(x), \hat{q}_\alpha(x_1), \hat{j}_\beta(x_2) \Big).\label{termtauc10}
\end{align}
Now, we write the correlation functions in the form of 
projectors and traces. We start by writing the most general 
expression for the three-point correlation function, involving 
three second-rank tensors and satisfying the orthogonality 
condition $u_\mu \hat{\pi}^{\mu\nu}$ and $u_\mu 
\hat{\tau}^{\mu\nu}$, given by
\begin{align}
	\Big( \hat{X}_{\mu\nu}(x), \hat{Y}_{\rho\sigma}(x_1),\hat{Z}_{\alpha\beta}(x_2) \Big)  &= a_1 \Delta_{\mu\nu}\Delta_{\rho\alpha}\Delta_{\sigma\beta} +a_2 \Delta_{\mu\nu}\Delta_{\rho\beta}\Delta_{\sigma\alpha} +b_1 \Delta_{\rho\sigma}\Delta_{\mu\alpha}\Delta_{\nu\beta} +b_2 \Delta_{\rho\sigma}\Delta_{\mu\beta}\Delta_{\nu\alpha}\nonumber\\
	&~~~ +c_1 \Delta_{\alpha\beta}\Delta_{\mu\rho}\Delta_{\nu\sigma} +c_2 \Delta_{\alpha\beta}\Delta_{\mu\sigma}\Delta_{\nu\rho} +d \Delta_{\mu\nu}\Delta_{\rho\sigma}\Delta_{\alpha\beta} \nonumber\\
	&~~~ +e_1 \Delta_{\mu\rho}\Delta_{\nu\alpha}\Delta_{\sigma\beta} +e_2 \Delta_{\mu\rho}\Delta_{\nu\beta}\Delta_{\sigma\alpha} +e_3 \Delta_{\mu\sigma}\Delta_{\nu\alpha}\Delta_{\rho\beta} +e_4 \Delta_{\mu\sigma}\Delta_{\nu\beta}\Delta_{\rho\alpha} \nonumber\\
	&~~~ +e_5 \Delta_{\mu\alpha}\Delta_{\nu\rho}\Delta_{\sigma\beta} +e_6 \Delta_{\mu\alpha}\Delta_{\nu\sigma}\Delta_{\rho\beta} +e_7 \Delta_{\mu\beta}\Delta_{\nu\rho}\Delta_{\sigma\alpha} +e_8 \Delta_{\mu\beta}\Delta_{\nu\sigma}\Delta_{\rho\alpha}, \label{general_six_decom}
\end{align}
where $\hat{X}$, $\hat{Y}$ and $\hat{Z}$ can either be $\hat{\pi}$ 
or $\hat{\tau}$. The coefficients $a_1$, $a_2$, $b_1$, $b_2$, 
$c_1$, $c_2$, $d$, $e_1$-$e_8$ are obtained by imposing the 
symmetric and antisymmetric properties of $\hat{\pi}_{\mu\nu}$ 
and $\hat{\tau}_{\mu\nu}$, respectively~(see 
Appendix~\ref{correlations_projectors}). Thus, we find that the 
three-point correlation functions among different two-rank 
tensors can be written as
\begin{align}
	\Big( \hat{\tau}_{\mu\nu}(x), \hat{\tau}_{\rho\sigma}(x_1),\hat{\tau}_{\alpha\beta}(x_2) \Big) &= \frac{1}{3} \Big[ \Delta_{\rho\alpha}\D_{\mu\nu\sigma\beta}-\Delta_{\rho\beta}\D_{\mu\nu\sigma\alpha}-\Delta_{\sigma\alpha}\D_{\mu\nu\rho\beta}+\Delta_{\sigma\beta}\D_{\mu\nu\rho\alpha} \Big] \Big(\hat{\tau}_\lambda^{~\delta}(x),\hat{\tau}_\delta^{~\eta}(x_1),\hat{\tau}_\eta^{~\lambda}(x_2) \Big),\label{cor1}\\
	\Big( \hat{\tau}_{\mu\nu}(x), \hat{\pi}_{\rho\sigma}(x_1),\hat{\pi}_{\alpha\beta}(x_2) \Big) &= \frac{-1}{15} \Big[ \Delta_{\rho\alpha}\D_{\mu\nu\sigma\beta}+\Delta_{\rho\beta}\D_{\mu\nu\sigma\alpha}+\Delta_{\sigma\alpha}\D_{\mu\nu\rho\beta}+\Delta_{\sigma\beta}\D_{\mu\nu\rho\alpha} \Big] \Big(\hat{\tau}_\lambda^{~\delta}(x),\hat{\pi}_\delta^{~\eta}(x_1),\hat{\pi}_\eta^{~\lambda}(x_2) \Big),\label{cor2}\\
	\Big( \hat{\tau}_{\mu\nu}(x), \hat{\pi}_{\rho\sigma}(x_1),\hat{\tau}_{\alpha\beta}(x_2) \Big) &= \frac{1}{5} \Big[ -\Delta_{\rho\alpha}\D_{\mu\nu\sigma\beta}+\Delta_{\rho\beta}\D_{\mu\nu\sigma\alpha}-\Delta_{\sigma\alpha}\D_{\mu\nu\rho\beta}+\Delta_{\sigma\beta}\D_{\mu\nu\rho\alpha}+\frac{4}{3}\Delta_{\rho\sigma}\D_{\mu\nu\alpha\beta} \Big]\nonumber\\
	&\qquad \times \Big(\hat{\tau}_\lambda^{~\delta}(x),\hat{\pi}_\delta^{~\eta}(x_1),\hat{\tau}_\eta^{~\lambda}(x_2) \Big),\label{cor3}\\
	\Big( \hat{\tau}_{\mu\nu}(x), \hat{\tau}_{\rho\sigma}(x_1),\hat{\pi}_{\alpha\beta}(x_2) \Big) &= \frac{-1}{5} \Big[ -\Delta_{\rho\alpha}\D_{\mu\nu\sigma\beta}-\Delta_{\rho\beta}\D_{\mu\nu\sigma\alpha}+\Delta_{\sigma\alpha}\D_{\mu\nu\rho\beta}+\Delta_{\sigma\beta}\D_{\mu\nu\rho\alpha}-\frac{4}{3}\Delta_{\alpha\beta}\D_{\mu\nu\rho\sigma} \Big]\nonumber\\
	&\qquad \times \Big(\hat{\tau}_\lambda^{~\delta}(x),\hat{\tau}_\delta^{~\eta}(x_1),\hat{\pi}_\eta^{~\lambda}(x_2) \Big).\label{cor4}
\end{align}
The other correlations including two second-rank tensors are 
given by (see Appendix~\ref{correlations_projectors})
\begin{align}
	\Big( \hat{\tau}_{\mu\nu}(x), \hat{\tau}_{\rho\sigma}(x_1),\hat{p}^*(x_2) \Big) &=\frac{1}{3}\D_{\mu\nu\rho\sigma}\Big( \hat{\tau}_{\delta\eta}(x), \hat{\tau}^{\delta\eta}(x_1),\hat{p}^*(x_2) \Big),\label{cor5}\\
	\Big( \hat{\tau}_{\mu\nu}(x),\hat{p}^*(x_1), \hat{\tau}_{\alpha\beta}(x_2) \Big) &=\frac{1}{3}\D_{\mu\nu\alpha\beta}\Big( \hat{\tau}_{\delta\eta}(x),\hat{p}^*(x_1),\hat{\tau}^{\delta\eta}(x_2) \Big),\label{cor6}\\
	\Big( \hat{\tau}_{\mu\nu}(x),\hat{q}_\alpha(x_1),\hat{q}_\beta(x_2) \Big) &=\frac{1}{3}\D_{\mu\nu\alpha\beta}\Big( \hat{\tau}_{\delta\eta}(x),\hat{q}^\delta(x_1),\hat{q}^\eta(x_2) \Big),\label{cor7}\\
	\Big( \hat{\tau}_{\mu\nu}(x),\hat{j}_\alpha(x_1),\hat{j}_\beta(x_2) \Big) &=\frac{1}{3}\D_{\mu\nu\alpha\beta}\Big( \hat{\tau}_{\delta\eta}(x),\hat{j}^\delta(x_1),\hat{j}^\eta(x_2) \Big)\label{cor8}\\
	\Big( \hat{\tau}_{\mu\nu}(x),\hat{j}_\alpha(x_1),\hat{q}_\beta(x_2) \Big) &=\frac{1}{3}\D_{\mu\nu\alpha\beta}\Big( \hat{\tau}_{\delta\eta}(x),\hat{j}^\delta(x_1),\hat{q}^\eta(x_2) \Big),\label{cor9}\\
	\Big( \hat{\tau}_{\mu\nu}(x),\hat{q}_\alpha(x_1),\hat{j}_\beta(x_2) \Big) &=\frac{1}{3}\D_{\mu\nu\alpha\beta}\Big( \hat{\tau}_{\delta\eta}(x),\hat{q}^\delta(x_1),\hat{j}^\eta(x_2) \Big).\label{cor10}
\end{align}
The symmetry property of the three-point correlation function 
indicates that $\mathcal{A}_{3\mu\nu}(x)=\mathcal{A}_{4\mu\nu}(x)$, $\mathcal{A}_
{5\mu\nu}(x)=\mathcal{A}_{6\mu\nu}(x)$, and 
$\mathcal{A}_{9\mu\nu}(x)=\mathcal{A}_{10\mu\nu}(x)$. 
Substituting the correlation functions 
from~\eqref{cor1}-\eqref{cor10} 
in~\eqref{termtauc1}-\eqref{termtauc10}, we obtain the second-order gradient correction from the three-point correlation function 
to the rotational stress tensor, given by
\begin{align}\label{finaltau3p}
	\big< \hat{\tau}_{\mu\nu}(x) \big>_2^{(3)} &= \gamma_{\tau\tau\tau}\xi_{\alpha\cll \mu }\xi_{\nu \crr}^{~\alpha}  +\gamma_{\tau\pi\pi}\sigma_{\alpha\cll \mu }\sigma_{\nu \crr}^{~\alpha} +2\gamma_{\tau\pi\tau}\sigma_{\alpha\cll \mu }\xi_{\nu \crr}^{~\alpha} +2\gamma_{\tau\tau p}\xi_{\mu\nu}\theta\nonumber\\
	&~~~ +\gamma_{\tau qq}M_{\cll \mu}M_{\nu\crr} +\gamma_{\tau jj}\beta^{-1}\nabla_{\cll \mu}\alpha \nabla_{\nu\crr}\alpha + 2\gamma_{\tau qj} M_{\cll\mu}\nabla_{\nu\crr}\alpha,
\end{align} 
where we used the shorthand notations~\eqref{shorthand3} and the 
new transport coefficients are defined by
\begin{align}
	\gamma_{\tau\tau\tau} &= \frac{-4}{3}\beta^2 \int d^4x_1 d^4x_2 \Big(\hat{\tau}_\lambda^{~\delta}(x),\hat{\tau}_\delta^{~\eta}(x_1),\hat{\tau}_\eta^{~\lambda}(x_2) \Big),\nonumber\\
	\gamma_{\tau\pi\pi} &= \frac{-4}{15}\beta^2 \int d^4x_1 d^4x_2 \Big(\hat{\tau}_\lambda^{~\delta}(x),\hat{\pi}_\delta^{~\eta}(x_1),\hat{\pi}_\eta^{~\lambda}(x_2) \Big),\nonumber\\
	\gamma_{\tau\pi\tau} &= \frac{4}{5}\beta^2 \int d^4x_1 d^4x_2 \Big(\hat{\tau}_\lambda^{~\delta}(x),\hat{\pi}_\delta^{~\eta}(x_1),\hat{\tau}_\eta^{~\lambda}(x_2) \Big),\nonumber\\
	\gamma_{\tau\tau p} &= \frac{-1}{3}\beta^2 \int d^4x_1 d^4x_2 \Big( \hat{\tau}_{\delta\eta}(x), \hat{\tau}^{\delta\eta}(x_1),\hat{p}^*(x_2) \Big),\nonumber\\
	\gamma_{\tau q q} &= \frac{1}{3}\beta^2  \int d^4x_1 d^4x_2 \Big( \hat{\tau}_{\delta\eta}(x), \hat{q}^{\delta}(x_1),\hat{q}^{\eta}(x_2) \Big),\nonumber\\
	\gamma_{\tau j j} &= \frac{1}{3} \beta \int d^4x_1 d^4x_2 \Big( \hat{\tau}_{\delta\eta}(x), \hat{j}^{\delta}(x_1),\hat{j}^{\eta}(x_2) \Big),\nonumber\\
	\gamma_{\tau q j} &= \frac{-1}{3} \beta \int d^4x_1 d^4x_2 \Big( \hat{\tau}_{\delta\eta}(x), \hat{q}^{\delta}(x_1),\hat{j}^{\eta}(x_2) \Big).\nonumber
\end{align}

\subsubsection{Final expression}

Using~\eqref{taufirst} and~\eqref{gamma_first} for the 
first-order correction 
and~\eqref{finaltau2nl},~\eqref{finaltau2et} 
and~\eqref{finaltau3p} for the second-order corrections, the 
complete expression for the rotational stress tensor $i.e.$ 
antisymmetric counterpart of the shear stress tensor, up to second 
order, is given by
\begin{align}
	\big< \hat{\tau}_{\mu\nu}(x) \big> 
	&= 2\gamma \xi_{\mu\nu} +2\gamma_\omega \left[ \D_{\mu\nu\rho\sigma} D\xi^{\rho\sigma} +\theta \Gamma \xi_{\mu\nu} \right] +  \gamma_{{}_{\tau S}} \mathcal{K}_{\mu\nu} \theta \nonumber\\
	&~~~ +\gamma_{\tau\tau\tau}\xi_{\alpha\cll \mu }\xi_{\nu \crr}^{~\alpha}  +\gamma_{\tau\pi\pi}\sigma_{\alpha\cll \mu }\sigma_{\nu \crr}^{~\alpha} +2\gamma_{\tau\pi\tau}\sigma_{\alpha\cll \mu }\xi_{\nu \crr}^{~\alpha} +2\gamma_{\tau\tau p}\xi_{\mu\nu}\theta\nonumber\\
	&~~~ +\gamma_{\tau qq}M_{\cll \mu}M_{\nu\crr} +\gamma_{\tau jj}\beta^{-1}\nabla_{\cll \mu}\alpha \nabla_{\nu\crr}\alpha + 2\gamma_{\tau qj} M_{\cll\mu}\nabla_{\nu\crr}\alpha .
\end{align}
Following~\cite{AmareshBhalerao, HARUTYUNYAN2022168755}, the evolution equation for the rotational stress tensor can be obtained by replacing $\xi^{\rho\sigma}\sim \frac{1}{2\gamma}\tau^{\rho\sigma}$ in the term $D\xi^{\rho\sigma}$, by employing~\eqref{transport}, the first-order Navier-Stokes relation. Thus, we obtain
\begin{equation}
	2\gamma_\omega \D_{\mu\nu\rho\sigma} D\xi^{\rho\sigma} \simeq -\tau_\tau \dot{\tau}_{\mu\nu}+\tilde{\gamma}_\tau \theta \tau_{\mu\nu},
\end{equation}
where,
\begin{align}
	\tau_\tau&=-\gamma_\omega \gamma^{-1},\\
	\tilde{\gamma}_\tau&=\tau_\tau \gamma^{-1}\beta \left( \Gamma\frac{\partial \gamma}{\partial \beta}-\delta\frac{\partial \gamma}{\partial \alpha} \right).
\end{align}
Therefore, the evolution equation for the rotational stress tensor is given by
\begin{align}\label{relaxtau}
	\tau_{\mu\nu}+\tau_\tau \dot{\tau}_{\mu\nu}
	&= 2\gamma \xi_{\mu\nu} +\tilde{\gamma}_\tau  \tau_{\mu\nu}\theta +2\gamma_\omega \Gamma\xi_{\mu\nu}\theta   +  \gamma_{{}_{\tau S}} \mathcal{K}_{\mu\nu} \theta \nonumber\\
	&~~~ +2\gamma_{\tau\tau p}\xi_{\mu\nu}\theta+\gamma_{\tau\tau\tau}\xi_{\alpha\cll \mu }\xi_{\nu \crr}^{~\alpha}  +\gamma_{\tau\pi\pi}\sigma_{\alpha\cll \mu }\sigma_{\nu \crr}^{~\alpha} +2\gamma_{\tau\pi\tau}\sigma_{\alpha\cll \mu }\xi_{\nu \crr}^{~\alpha} \nonumber\\
	&~~~ +\gamma_{\tau qq}M_{\cll \mu}M_{\nu\crr} +\gamma_{\tau jj}\beta^{-1}\nabla_{\cll \mu}\alpha \nabla_{\nu\crr}\alpha + 2\gamma_{\tau qj} M_{\cll\mu}\nabla_{\nu\crr}\alpha .
\end{align}

\subsection{Boost Heat Vector ($q_\mu$)}

\subsubsection{Nonlocal correction from two-point correlation function: $\big< \hat{q}_\mu(x) \big>_2^{(2),NL}$}
Invoking the Curie's principle for the isotropic medium and 
using~\eqref{nlc} and~\eqref{CF}, we have
\begin{equation}\label{q1}
	\big< \hat{q}_\mu(x) \big>_2^{(2),NL}	= \int d^4 x_1 \Big( \hat{q}_\mu(x), \hat{q}_\nu(x_1) \Big)\beta(x_1)M^\nu(x_1)-\big< \hat{q}_\mu(x) \big>_1 .
\end{equation}
In an isotropic medium, the correlator can be written as
\begin{equation}\label{q_corr}
	\Big( \hat{q}_\mu(x), \hat{q}_\nu(x_1) \Big) = \frac{1}{3} \Delta_{\mu\nu} \Big( \hat{q}^\sigma (x), \hat{q}_\sigma(x_1) \Big).
\end{equation}
Similar to the calculation of a nonlocal correction to $\big< 
\hat{\tau}_{\mu\nu}(x) \big>_2$, we have to first express 
$\hat{q}^\mu$ in terms of the energy-momentum tensor, and then we 
expand all the hydrodynamic quantities around $x_1=x$. 
From~\eqref{projectionsTN}, in the Landau frame, we can write
\begin{equation}
	\hat{q}_\sigma (x_1) = \Delta_{\sigma\alpha} (x_1) u_\beta (x_1) \hat{T}^{\alpha\beta}(x_1).
\end{equation}
Using the Taylor expansion, we expand all the hydrodynamic 
quantities $\Delta_{\sigma\alpha} (x_1)$ and $u_\beta (x_1)$ (not 
operators) around $x_1=x$, and keeping terms only up to 
first-order in gradient expansion, we have
\begin{equation}\label{q_tay_expn}
	\hat{q}_\sigma(x_1) = \left[\hat{q}_\sigma(x_1)\right]_x + (x_1-x)^\tau \left[\frac{\partial \hat{q}_\sigma(x_1)}{\partial x_1^\tau}\right]_x ,
\end{equation}
where, 
\begin{align}
	\left[\hat{q}(x_1)\right]_x &= \Delta_{\sigma\alpha} (x) u_\beta (x) \hat{T}^{\alpha\beta}(x_1) ,\\
	\left[\frac{\partial \hat{q}_\sigma(x_1)}{\partial x_1^\tau}\right]_x &= -\hat{q}^\alpha(x_1) u_\sigma(x) \frac{\partial u_\alpha(x_1)}{\partial x_1^\tau}\bigg|_x - \hat{\varepsilon}(x_1) \frac{\partial u_\sigma(x_1)}{\partial x_1^\tau}\bigg|_x-\hat{p}(x_1) \frac{\partial u_\sigma(x_1)}{\partial x_1^\tau}\bigg|_x+ [\hat{\pi}_\sigma^{~\beta}(x_1)+\hat{\tau}_\sigma^{~\beta}(x_1)]\frac{\partial u_\beta(x_1)}{\partial x_1^\tau}\bigg|_x ,\label{delq}
\end{align}
defining $M'^\lambda=\beta M^\lambda$, using the Taylor expansion 
for $M'^\lambda (x_1)$, given by
\begin{equation}\label{M_tay_expn}
	M'^\lambda (x_1) = M'^\lambda (x) +(x_1-x)^\tau \frac{\partial M'^\lambda (x_1)}{\partial x_1^\tau}\bigg|_x +\cdots.
\end{equation}
Substituting~\eqref{q_corr},~\eqref{q_tay_expn}, 
and~\eqref{M_tay_expn} in~\eqref{q1} and, collecting all the 
terms up to second-order in gradient expansion, we obtain
\begin{align}
	\big< \hat{q}_\mu(x) \big>_2^{(2),NL}	&= \frac{1}{3}\Delta_{\mu\lambda}(x)\int d^4 x_1 \Bigg[ \Big( \hat{q}^\sigma(x),\left[\hat{q}_\sigma(x_1)\right]_x \Big)M'^\lambda(x) + \Big( \hat{q}^\sigma(x),\left[\hat{q}_\sigma(x_1)\right]_x \Big)(x_1-x)^\tau \frac{\partial M'^\lambda (x_1)}{\partial x_1^\tau}\bigg|_x \nonumber\\
	& \hspace{3.2cm} + \left(\hat{q}^\sigma(x),\left[\frac{\partial q_\sigma (x_1)}{\partial x_1^\tau}\right]_x \right)(x_1-x)^\tau M'^\lambda(x)  \Bigg] -\big< \hat{q}_\mu(x) \big>_1.
\end{align}
With the help of~\eqref{qfirst}, the first term in the bracket 
cancels $\big< \hat{q}_\mu(x) \big>_1$. Thus, we obtain
\begin{align}\label{q_term2}
	\big< \hat{q}_\mu(x) \big>_2^{(2),NL}	&= \frac{1}{3}\Delta_{\mu\lambda}(x) \partial_\tau M'^\lambda(x) \int d^4 x_1 \Big( \hat{q}^\sigma(x),\left[\hat{q}_\sigma(x_1)\right]_x \Big)(x_1-x)^\tau \nonumber\\
	&~~~ +\frac{1}{3} M'_\mu(x) \int d^4 x_1 \bigg(\hat{q}^\sigma(x),\left[\frac{\partial q_\sigma (x_1)}{\partial x_1^\tau}\right]_x \bigg)(x_1-x)^\tau.
\end{align}
Observing~\eqref{delq}, we can conclude that the second term 
in~\eqref{q_term2} vanishes, on account of Curie's principle and 
the orthogonal property $\hat{q}^\sigma u_\sigma=0$. 
Thus,~\eqref{q_term2} can be simplified to
\begin{equation}\label{q_term3}
	\big< \hat{q}_\mu(x) \big>_2^{(2),NL} = [\beta(x)]^{-1}\Delta_{\mu\lambda}(x) R^\tau(x)\partial_\tau M'^\lambda(x) ,
\end{equation}
where we have approximated $\left[\hat{q}(x_1)\right]_x \simeq 
\hat{q}(x_1)$, because~\eqref{q_term3} already contains a 
second-order term. We also defined
\begin{equation}
	R^\tau(x) = \frac{\beta(x)}{3} \int d^4 x_1 \Big( \hat{q}^\sigma(x),\hat{q}_\sigma(x_1) \Big)(x_1-x)^\tau,
\end{equation}
which can be simplified in the following form (see 
Eq.~\eqref{freq_dep_trans} in Appendix~\ref{correlators})
\begin{equation}\label{def_transq_a2}
	R^\tau = \lambda_\omega u^\tau,
\end{equation}
where
\begin{equation}
	\lambda_\omega =  i\lim_{\omega\to 0}\frac{d}{d\omega}\lambda_{\hat{q}\hat{q}}(\omega).
\end{equation}
Here $\lambda_{\hat{q}\hat{q}}(\omega)$ is defined by (see 
Appendix~\ref{correlators})
\begin{eqnarray}
	\lambda_{\hat{q}\hat{q}}(\omega)=\frac{\beta}{3} \int d^4x_1 e^{i\omega(t-t_1)}\Big( \hat{q}^{\sigma}(x), \hat{q}_{\sigma}(x_1) \Big).
\end{eqnarray}
Finally, solving~\eqref{q_term3} using~\eqref{def_transq_a2}, and 
using the first-order approximation 
$D\beta\simeq\Gamma\beta\theta$, we obtain
\begin{equation}\label{q_final_2nl}
	\big< \hat{q}_\mu \big>_2^{(2),NL} = \lambda_\omega [ M_\mu \Gamma \theta +\Delta_{\mu\lambda}D M^\lambda ].
\end{equation}

\subsubsection{Local correction from two-point correlation function using extended thermodynamic forces: $\big< \hat{q}_\mu(x)\big>_2^{(2),ET}$}
From~\eqref{etc}, we can write
\begin{equation}\label{q_et_def}
	\big< \hat{q}_\mu(x)\big>_2^{(2),ET} = \int d^4 x_1 \Big( \hat{q}_\mu(x), \hat{C}_S(x_1) \Big)\Big|_{local},
\end{equation}
where $\hat{C}_S$ is given by~\eqref{CS}. $\hat{C}_S$ has only 
one vector term, then by the help of Curie's principle, and 
expanding all hydrodynamic quantities around $x_1=x$, we can 
extract the local terms in~\eqref{q_et_def}, given by
\begin{equation}\label{q_final_2et}
	\big< \hat{q}_{\mu}(x) \big>_2^{(2),ET} = -\lambda(x)	\mathcal{Q}_\mu(x),
\end{equation}
where, 
\begin{equation}
	\lambda(x) = \frac{-\beta}{3} \int d^4 x_1 \Big( \hat{q}^\alpha (x), \hat{q}_\alpha (x_1) \Big).
\end{equation}

\subsubsection{Local correction from three-point correlation function: $\big< \hat{q}_{\mu}(x) \big>_2^{(3)}$}

From~\eqref{tpc}, we can write
\begin{equation}\label{q_3p_def}
	\big< \hat{q}_\mu(x) \big>_2^{(3)}	= \int d^4 x_1 d^4 x_2 \Big( \hat{q}_\mu(x), \hat{C}_F(x_1),\hat{C}_F(x_2) \Big) \Big|_{local},
\end{equation}
where, $\hat{C}_F$ is given by~\eqref{CF}. 
Substituting~\eqref{CF} in~\eqref{q_3p_def} and employing Curie's 
principle, the nonvanishing correlations are given by
\begin{align}
	\mathcal{B}_{1\mu}(x) &= -\beta^2(x)M^\nu(x)\theta(x)\int d^4 x_1 d^4 x_2 \Big( \hat{q}_\mu(x), \hat{q}_\nu(x_1),\hat{p}^*(x_2) \Big),  \label{termqc1}\\
	\mathcal{B}_{2\mu}(x) &= -\beta^2(x)M^\nu(x)\theta(x)\int d^4 x_1 d^4 x_2 \Big( \hat{q}_\mu(x), \hat{p}^*(x_1), \hat{q}_\nu(x_2) \Big),  \label{termqc2}\\
	\mathcal{B}_{3\mu}(x) &= \beta^2(x)M^\nu(x)\xi^{\alpha\beta}(x)\int d^4 x_1 d^4 x_2 \Big( \hat{q}_\mu(x), \hat{q}_\nu(x_1),\hat{\tau}_{\alpha\beta}(x_2) \Big) , \label{termqc3}\\
	\mathcal{B}_{4\mu}(x) &= \beta^2(x)M^\nu(x)\xi^{\rho\sigma}(x)\int d^4 x_1 d^4 x_2 \Big( \hat{q}_\mu(x), \hat{\tau}_{\rho\sigma}(x_1),\hat{q}_\nu(x_2), \Big) , \label{termqc4}\\
	\mathcal{B}_{5\mu}(x) &= \beta^2(x)M^\nu(x)\sigma^{\alpha\beta}(x)\int d^4 x_1 d^4 x_2 \Big( \hat{q}_\mu(x), \hat{q}_\nu(x_1),\hat{\pi}_{\alpha\beta}(x_2) \Big) , \label{termqc5}\\
	\mathcal{B}_{6\mu}(x) &= \beta^2(x)M^\nu(x)\sigma^{\rho\sigma}(x)\int d^4 x_1 d^4 x_2 \Big( \hat{q}_\mu(x), \hat{\pi}_{\rho\sigma}(x_1),\hat{q}_\nu(x_2), \Big) , \label{termqc6}\\
	\mathcal{B}_{7\mu}(x) &= -\beta(x)\nabla^\nu\alpha(x)\xi^{\alpha\beta}(x)\int d^4 x_1 d^4 x_2 \Big( \hat{q}_\mu(x), \hat{j}_\nu(x_1),\hat{\tau}_{\alpha\beta}(x_2) \Big) , \label{termqc7}\\
	\mathcal{B}_{8\mu}(x) &= -\beta(x)\nabla^\nu\alpha(x)\xi^{\alpha\beta}(x)\int d^4 x_1 d^4 x_2 \Big( \hat{q}_\mu(x),\hat{\tau}_{\alpha\beta}(x_1), \hat{j}_\nu(x_2) \Big),  \label{termqc8}\\
	\mathcal{B}_{9\mu}(x) &= -\beta(x)\nabla^\nu\alpha(x)\sigma^{\alpha\beta}(x)\int d^4 x_1 d^4 x_2 \Big( \hat{q}_\mu(x), \hat{j}_\nu(x_1),\hat{\pi}_{\alpha\beta}(x_2) \Big) , \label{termqc9}\\
	\mathcal{B}_{10\mu}(x) &= -\beta(x)\nabla^\nu\alpha(x)\sigma^{\alpha\beta}(x)\int d^4 x_1 d^4 x_2 \Big( \hat{q}_\mu(x),\hat{\pi}_{\alpha\beta}(x_1), \hat{j}_\nu(x_2) \Big).  \label{termqc10}
\end{align}
Again the symmetric property of the three-point function leads to 
$\mathcal{B}_{1\mu}(x)=\mathcal{B}_{2\mu}(x)$, 
$\mathcal{B}_{3\mu}(x)=\mathcal{B}_{4\mu}(x)$, 
$\mathcal{B}_{5\mu}(x)=\mathcal{B}_{6\mu}(x)$, 
$\mathcal{B}_{7\mu}(x)=\mathcal{B}_{8\mu}(x)$, and 
$\mathcal{B}_{9\mu}(x)=\mathcal{B}_{10\mu}(x)$. In the isotopic 
medium, the three-point correlation functions can be written as
\begin{align}
	\Big( \hat{q}_\mu(x), \hat{q}_\nu(x_1),\hat{p}^*(x_2) \Big) &= \frac{1}{3} \Delta_{\mu\nu} \Big( \hat{q}_\lambda(x), \hat{q}^\lambda(x_1),\hat{p}^*(x_2) \Big),\label{termcorrq1}\\
	\Big( \hat{q}_\mu(x), \hat{q}_\nu(x_1),\hat{\tau}_{\alpha\beta}(x_2) \Big)  &= \frac{1}{3} \D_{\mu\nu\alpha\beta} \Big( \hat{q}^\lambda(x), \hat{q}^\delta(x_1), \hat{\tau}_{\lambda\delta}(x_2) \Big),\label{termcorrq2}\\
	\Big( \hat{q}_\mu(x), \hat{q}_\nu(x_1),\hat{\pi}_{\alpha\beta}(x_2) \Big)  &= \frac{1}{5} \Delta_{\mu\nu\alpha\beta} \Big( \hat{q}^\lambda(x), \hat{q}^\delta(x_1), \hat{\pi}_{\lambda\delta}(x_2) \Big),\label{termcorrq3}\\
	\Big( \hat{q}_\mu(x), \hat{j}_\nu(x_1),\hat{\tau}_{\alpha\beta}(x_2) \Big)  &= \frac{1}{3} \D_{\mu\nu\alpha\beta} \Big( \hat{q}^\lambda(x), \hat{j}^\delta(x_1), \hat{\tau}_{\lambda\delta}(x_2) \Big),\label{termcorrq4}\\
	\Big( \hat{q}_\mu(x), \hat{j}_\nu(x_1),\hat{\pi}_{\alpha\beta}(x_2) \Big)  &= \frac{1}{5} \Delta_{\mu\nu\alpha\beta} \Big( \hat{q}^\lambda(x), \hat{j}^\delta(x_1), \hat{\pi}_{\lambda\delta}(x_2) \Big).\label{termcorrq5}
\end{align}
Substituting~\eqref{termcorrq1}-\eqref{termcorrq5} 
in~\eqref{termqc1}-\eqref{termqc10}, we get the final expression 
for the second-order correction from the three-point correlation 
function in $\big<\hat{q}_\mu\big>$, given by
\begin{equation}\label{q_final_3p}
	\big< \hat{q}_\mu \big>_2^{(3)} = \lambda_{qqp} M_\mu\theta + \lambda_{qq\tau} \xi_{\mu\nu}M^\nu +\lambda_{qq\pi} \sigma_{\mu\nu}M^\nu +\lambda_{qj\tau} \xi_{\mu\nu}\nabla^\nu\alpha +\lambda_{qj\pi} \sigma_{\mu\nu}\nabla^\nu\alpha,
\end{equation}
where, the new transport coefficients are defined by
\begin{align}
	\lambda_{qqp} &= \frac{-2}{3} \beta^2  \int d^4x_1 d^4x_2 \Big( \hat{q}_\lambda(x), \hat{q}^\lambda(x_1),\hat{p}^*(x_2) \Big),\\
	\lambda_{qq\tau} &= \frac{2}{3} \beta^2 \int d^4x_1 d^4x_2 \Big( \hat{q}^\lambda(x), \hat{q}^\delta(x_1), \hat{\tau}_{\lambda\delta}(x_2) \Big), \\
	\lambda_{qq\pi} &= \frac{2}{5} \beta^2 \int d^4x_1 d^4x_2 \Big( \hat{q}^\lambda(x), \hat{q}^\delta(x_1), \hat{\pi}_{\lambda\delta}(x_2) \Big),\\
	\lambda_{qj\tau} &= -\frac{2}{3} \beta \int d^4x_1 d^4x_2 \Big( \hat{q}^\lambda(x), \hat{j}^\delta(x_1), \hat{\tau}_{\lambda\delta}(x_2) \Big),\\
	\lambda_{qj\pi} &= -\frac{2}{5} \beta \int d^4x_1 d^4x_2 \Big( \hat{q}^\lambda(x), \hat{j}^\delta(x_1), \hat{\pi}_{\lambda\delta}(x_2) \Big).
\end{align}

\subsubsection{Final expression}
Combining~\eqref{qfirst},~\eqref{lambda_first},~\eqref{q_final_2nl},
~\eqref{q_final_2et}, and~\eqref{q_final_3p}, the complete 
second-order expression for boost heat vector, $i.e.$, the antisymmetric counterpart of the heat flow vector, is given by
\begin{align}
	\big< \hat{q}_\mu \big> &= -\lambda M_\mu +\lambda_\omega [ M_\mu \Gamma \theta +\Delta_{\mu\lambda}D M^\lambda ]-\lambda \mathcal{Q}_\mu\nonumber\\
	&~~~ +\lambda_{qqp} M_\mu\theta + \lambda_{qq\tau} \xi_{\mu\nu}M^\nu +\lambda_{qq\pi} \sigma_{\mu\nu}M^\nu +\lambda_{qj\tau} \xi_{\mu\nu}\nabla^\nu\alpha +\lambda_{qj\pi} \sigma_{\mu\nu}\nabla^\nu\alpha.
\end{align}
Approximating, $M^\lambda\sim \lambda^{-1} q^\lambda$, by using the first-order Navier-Stokes relation~\eqref{transport}, in the term $D M^\lambda$. We get the evolution equation for the boost heat vector
\begin{align}\label{relaxq}
	q_\mu +\tau_q \dot{q}_\mu&= -\lambda M_\mu +\tilde{\lambda}_\tau \theta q_\mu+\lambda_\omega \Gamma M_\mu\theta  -\lambda \mathcal{Q}_\mu\nonumber\\
	&~~~  +\lambda_{qqp} M_\mu\theta+ \lambda_{qq\tau} \xi_{\mu\nu}M^\nu +\lambda_{qq\pi} \sigma_{\mu\nu}M^\nu +\lambda_{qj\tau} \xi_{\mu\nu}\nabla^\nu\alpha +\lambda_{qj\pi} \sigma_{\mu\nu}\nabla^\nu\alpha.
\end{align}
where,
\begin{align}
	\tau_q&=\lambda_\omega \lambda^{-1}\\
	\tilde{\lambda}_\tau&=\tau_q \lambda^{-1}\beta \left( \Gamma\frac{\partial \lambda}{\partial \beta}-\delta\frac{\partial \lambda}{\partial \alpha} \right)
\end{align}

\subsection{Shear Stress Tensor ($\pi_{\mu\nu}$)}

\subsubsection{Nonlocal correction from two-point correlation function: $\big< \hat{\pi}_{\mu\nu}(x) \big>_2^{(2),NL}$}
Invoking Curie's principle and using~\eqref{nlc} with~\eqref{CF}, 
we can write
\begin{equation}
	\big< \hat{\pi}_{\mu\nu}(x) \big>_2^{(2),NL} =  \int d^4 x_1 \Big( \hat{\pi}_{\mu\nu}(x), \hat{\pi}_{\alpha\beta}(x_1) \Big)\beta(x_1)\sigma^{\alpha\beta}(x_1)-\big< \hat{\pi}_{\mu\nu}(x) \big>_1 . \label{pinlstart}
\end{equation}
In this case, the nonlocal generalization of the two-point correlation function is given by
\begin{equation}
	\Big( \hat{\pi}_{\mu\nu}(x), \hat{\pi}_{\alpha\beta}(x_1) \Big) = \frac{1}{5}\Delta_{\mu\nu\alpha\beta }(x,x_1)\Big( \hat{\pi}^{\lambda\eta}(x),\hat{\pi}_{\lambda\eta}(x_1) \Big).\label{picorrdecom}
\end{equation}
Next, we write $\hat{\pi}^{\lambda\eta}$ and 
$\hat{\pi}_{\lambda\eta}$ in terms of energy-momentum tensor. 
Using~\eqref{projectionsTN}, we have
\begin{align}
	\hat{\pi}^{\lambda\eta}(x) &= \Delta_{\gamma\delta}^{\lambda\eta}(x)\hat{T}^{\gamma\delta}(x), \label{piupproj}\\
	\hat{\pi}_{\lambda\eta}(x_1) &= \Delta_{\lambda\eta\alpha\beta}(x_1)\hat{T}^{\alpha\beta}(x_1).\label{pidownproj}
\end{align}
Substituting~\eqref{picorrdecom} with~\eqref{piupproj} 
and~\eqref{pidownproj} in~\eqref{pinlstart}, we obtain
\begin{equation}
	\big< \hat{\pi}_{\mu\nu}(x) \big>_2^{(2),NL}= \frac{1}{5}\Delta_{\mu\nu\rho\sigma }(x) \int d^4 x_1 \Big( \hat{T}^{\gamma\delta}(x), \hat{T}^{\alpha\beta}(x_1) \Big)\Delta_{\gamma\delta\alpha\beta}(x,x_1)\beta(x_1)\sigma^{\rho\sigma}(x_1)-\big< \hat{\pi}_{\mu\nu}(x) \big>_1 . \label{pi_nl_term2}
\end{equation}
The procedure for the calculation of nonlocal correction from 
two-point correlation function to $\big< \hat{\pi}_{\mu\nu}(x) 
\big>_2$ is similar to that of $\big< \hat{\tau}_{\mu\nu}(x) 
\big>_2$, with the antisymmetric fourth-rank projector replaced 
by the symmetric fourth-rank projector. Hence, as discussed 
in Sec.~\ref{tau_nlc}, expanding all the hydrodynamic quantities 
around $x_1=x$ in~\eqref{pi_nl_term2}, then collecting all the 
second-order terms, we obtain
\begin{equation}
	\big< \hat{\pi}_{\mu\nu}(x) \big>_2^{(2),NL} = 2 \Delta_{\mu\nu\rho\sigma}(x)[\beta(s)]^{-1}K^\lambda(x)\frac{\partial }{\partial x_1^\lambda}[\beta(x_1)\sigma^{\rho\sigma}(x_1)]\bigg|_{x_1=x}. \label{pi_nl_term4}
\end{equation}
Here we defined
\begin{equation}
	K^\lambda(x) =  \frac{\beta(x)}{10}\int d^4 x_1 \Big( \hat{\pi}_{\alpha\beta}(x),\hat{\pi}^{\alpha\beta}(x_1) \Big)(x_1-x)^\lambda,
\end{equation}
which can be written as (see Eq.~\eqref{freq_dep_trans} in 
Appendix~\ref{correlators})
\begin{equation}\label{def_transpi_a2}
	K^\lambda = \eta_\omega u^\lambda,
\end{equation}
where, 
\begin{equation}
	\eta_\omega = i\lim_{\omega\to 0}\frac{d}{d\omega}\eta_{\hat{\pi}\hat{\pi}}(\omega).
\end{equation}
The frequency-dependent transport coefficient 
$\eta_{\hat{\pi}\hat{\pi}}(\omega)$ is given by (see 
Appendix~\ref{correlators})
\begin{equation}
	\eta_{\hat{\pi}\hat{\pi}}(\omega)=\frac{\beta}{10} \int d^4x_1 e^{i\omega(t-t_1)}\Big( \hat{\pi}^{\alpha\beta}(x), \hat{\pi}_{\alpha\beta}(x_1) \Big).
\end{equation}
Solving~\eqref{pi_nl_term4} with~\eqref{def_transpi_a2} and using 
$D\beta\simeq\Gamma\beta\theta$, we obtain
\begin{equation}\label{finalpi2nl}
	\big< \hat{\pi}_{\mu\nu}\big>_2^{(2),NL} = 2\eta_\omega \left[ \Delta_{\mu\nu\rho\sigma} D\sigma^{\rho\sigma} +\theta \Gamma \sigma_{\mu\nu} \right].
\end{equation}

\subsubsection{Local correction from two-point correlation function using extended thermodynamic forces: $\big< \hat{\pi}_{\mu\nu}(x) \big>_2^{(2),ET}$}
With the help of~\eqref{etc}, we can write
\begin{equation}
	\left< \hat{\pi}_{\mu\nu}(x)\right>_2^{(2),ET} = \int d^4 x_1 \Big( \hat{\pi}_{\mu\nu}(x), \hat{C}_S(x_1) \Big)\Big|_{local},
\end{equation}
where $\hat{C}_S$ is given by~\eqref{CS}. Since~\eqref{CS} does 
not contain any term involving a symmetric second-rank operator, 
hence all the correlations vanish by employing Curie's 
principle. Therefore,
\begin{equation}\label{finalpi2et}
	\left< \hat{\pi}_{\mu\nu}(x)\right>_2^{(2),ET} = 0.
\end{equation}

\subsubsection{Local correction from three-point correlation function: $\big< \hat{\pi}_{\mu\nu}(x) \big>_2^{(3)}$}

From~\eqref{tpc}, we can write
\begin{equation}
	\big< \hat{\pi}_{\mu\nu}(x) \big>_2^{(3)}	= \int d^4 x_1 d^4 x_2 \Big( \hat{\pi}_{\mu\nu}(x), \hat{C}_F(x_1),\hat{C}_F(x_2) \Big) \Big|_{local},\label{pi_3p_start}
\end{equation}
where $\hat{C}_F$ is given by~\eqref{CF}. Here, we follow the 
procedure given in Sec.~\ref{tau3p}. Substituting~\eqref{CF} 
in~\eqref{pi_3p_start}, the nonvanishing correlations, involving 
three second-rank tensor operators are given by (see 
Appendix~\ref{correlations_projectors})
\begin{align}
	\Big( \hat{\pi}_{\mu\nu}(x), \hat{\pi}_{\rho\sigma}(x_1),\hat{\pi}_{\alpha\beta}(x_2) \Big) &=   \frac{1}{35} \Big[ 3(\Delta_{\rho\alpha}\Delta_{\mu\nu\sigma\beta}+\Delta_{\rho\beta}\Delta_{\mu\nu\sigma\alpha}+\Delta_{\sigma\alpha}\Delta_{\mu\nu\rho\beta}+\Delta_{\sigma\beta}\Delta_{\mu\nu\rho\alpha})\nonumber\\ &\quad-4(\Delta_{\rho\sigma}\Delta_{\mu\nu\alpha\beta}+\Delta_{\alpha\beta}\Delta_{\mu\nu\rho\sigma})\Big]\Big(\hat{\pi}_\lambda^{~\delta}(x),\hat{\pi}_\delta^{~\eta}(x_1),\hat{\pi}_\eta^{~\lambda}(x_2) \Big),\label{picor1}\\
	\Big( \hat{\pi}_{\mu\nu}(x), \hat{\tau}_{\rho\sigma}(x_1),\hat{\tau}_{\alpha\beta}(x_2) \Big) &= \frac{-1}{5} \Big[ \Delta_{\rho\alpha}\Delta_{\mu\nu\sigma\beta}-\Delta_{\rho\beta}\Delta_{\mu\nu\sigma\alpha}-\Delta_{\sigma\alpha}\Delta_{\mu\nu\rho\beta}+\Delta_{\sigma\beta}\Delta_{\mu\nu\rho\alpha} \Big]\nonumber\\
	&\quad\times  \Big(\hat{\pi}_\lambda^{~\delta}(x),\hat{\tau}_\delta^{~\eta}(x_1),\hat{\tau}_\eta^{~\lambda}(x_2) \Big),\label{picor2}\\
	\Big( \hat{\pi}_{\mu\nu}(x), \hat{\pi}_{\rho\sigma}(x_1),\hat{\tau}_{\alpha\beta}(x_2) \Big) &= \frac{-1}{15} \Big[ -\Delta_{\rho\alpha}\Delta_{\mu\nu\sigma\beta}+\Delta_{\rho\beta}\Delta_{\mu\nu\sigma\alpha}-\Delta_{\sigma\alpha}\Delta_{\mu\nu\rho\beta}+\Delta_{\sigma\beta}\Delta_{\mu\nu\rho\alpha} \Big]\nonumber\\
	&\quad\times  \Big(\hat{\pi}_\lambda^{~\delta}(x),\hat{\pi}_\delta^{~\eta}(x_1),\hat{\tau}_\eta^{~\lambda}(x_2) \Big),\label{picor3}\\
	\Big( \hat{\pi}_{\mu\nu}(x), \hat{\tau}_{\rho\sigma}(x_1),\hat{\pi}_{\alpha\beta}(x_2) \Big) &= \frac{1}{15} \Big[ -\Delta_{\rho\alpha}\Delta_{\mu\nu\sigma\beta}-\Delta_{\rho\beta}\Delta_{\mu\nu\sigma\alpha}+\Delta_{\sigma\alpha}\Delta_{\mu\nu\rho\beta}+\Delta_{\sigma\beta}\Delta_{\mu\nu\rho\alpha} \Big]\nonumber\\
	&\quad\times  \Big(\hat{\pi}_\lambda^{~\delta}(x),\hat{\tau}_\delta^{~\eta}(x_1),\hat{\pi}_\eta^{~\lambda}(x_2) \Big),\label{picor4}
\end{align}
whereas, other nonvanishing correlations are given by
\begin{align}	
	\Big( \hat{\pi}_{\mu\nu}(x), \hat{\pi}_{\rho\sigma}(x_1),\hat{p}^*(x_2) \Big) &=\frac{1}{5}\Delta_{\mu\nu\rho\sigma}\Big( \hat{\pi}_{\delta\eta}(x), \hat{\pi}^{\delta\eta}(x_1),\hat{p}^*(x_2) \Big),\label{picor5}\\
	\Big( \hat{\pi}_{\mu\nu}(x),\hat{p}^*(x_1), \hat{\pi}_{\alpha\beta}(x_2) \Big) &=\frac{1}{5}\Delta_{\mu\nu\alpha\beta}\Big( \hat{\pi}_{\delta\eta}(x),\hat{p}^*(x_1),\hat{\pi}^{\delta\eta}(x_2) \Big),\label{picor6}\\
	\Big( \hat{\pi}_{\mu\nu}(x),\hat{q}_\alpha(x_1),\hat{q}_\beta(x_2) \Big) &=\frac{1}{5}\Delta_{\mu\nu\alpha\beta}\Big( \hat{\pi}_{\delta\eta}(x),\hat{q}^\delta(x_1),\hat{q}^\eta(x_2) \Big),\label{picor7}\\
	\Big( \hat{\pi}_{\mu\nu}(x),\hat{j}_\alpha(x_1),\hat{j}_\beta(x_2) \Big) &=\frac{1}{5}\Delta_{\mu\nu\alpha\beta}\Big( \hat{\pi}_{\delta\eta}(x),\hat{j}^\delta(x_1),\hat{j}^\eta(x_2) \Big),\label{picor8}\\
	\Big( \hat{\pi}_{\mu\nu}(x),\hat{j}_\alpha(x_1),\hat{q}_\beta(x_2) \Big) &=\frac{1}{5}\Delta_{\mu\nu\alpha\beta}\Big( \hat{\pi}_{\delta\eta}(x),\hat{j}^\delta(x_1),\hat{q}^\eta(x_2) \Big),\label{picor9}\\
	\Big( \hat{\pi}_{\mu\nu}(x),\hat{q}_\alpha(x_1),\hat{j}_\beta(x_2) \Big) &=\frac{1}{5}\Delta_{\mu\nu\alpha\beta}\Big( \hat{\pi}_{\delta\eta}(x),\hat{q}^\delta(x_1),\hat{j}^\eta(x_2) \Big).\label{picor10}
\end{align}
Finally, with the help of correlation functions given 
in~\eqref{picor1}-\eqref{picor10}, approximating all the 
thermodynamics forces at $x$, we obtain  
\begin{align}\label{finalpi3p}
	\big< \hat{\pi}_{\mu\nu}(x) \big>_2^{(3)}	 &= \eta_{\pi\pi\pi} \sigma_{\alpha <\mu}\sigma_{\nu>}^{~\alpha} + \eta_{\pi\tau\tau} \xi_{\alpha <\mu}\xi_{\nu>}^{~\alpha}+2\eta_{\pi\pi\tau} \sigma_{\alpha <\mu}\xi_{\nu>}^{~\alpha}+ 2\eta_{\pi\pi p} \sigma_{\mu\nu}\theta \nonumber\\
	&~~~ +\eta_{\pi qq}M_{< \mu}M_{\nu>} +\eta_{\pi jj}\beta^{-1}\nabla_{< \mu}\alpha \nabla_{\nu>}\alpha + 2\eta_{\pi qj} M_{<\mu}\nabla_{\nu>}\alpha,
\end{align}
where, the various transport coefficients are defined by
\begin{align}
	\eta_{\pi\pi\pi} &= \frac{12}{35}\beta^2 \int d^4x_1 d^4x_2 \Big(\hat{\pi}_\lambda^{~\delta}(x),\hat{\pi}_\delta^{~\eta}(x_1),\hat{\pi}_\eta^{~\lambda}(x_2) \Big),\nonumber\\
	\eta_{\pi\tau\tau} &= \frac{4}{5}\beta^2 \int d^4x_1 d^4x_2 \Big(\hat{\pi}_\lambda^{~\delta}(x),\hat{\tau}_\delta^{~\eta}(x_1),\hat{\tau}_\eta^{~\lambda}(x_2) \Big),\nonumber\\
	\eta_{\pi\pi\tau} &= \frac{-4}{15}\beta^2 \int d^4x_1 d^4x_2 \Big(\hat{\pi}_\lambda^{~\delta}(x),\hat{\pi}_\delta^{~\eta}(x_1),\hat{\tau}_\eta^{~\lambda}(x_2) \Big),\nonumber\\
	\eta_{\pi\pi p} &= \frac{-1}{5}\beta^2 \int d^4x_1 d^4x_2 \Big( \hat{\pi}_{\delta\eta}(x), \hat{\pi}^{\delta\eta}(x_1),\hat{p}^*(x_2) \Big),\nonumber\\
	\eta_{\pi q q} &= \frac{1}{5}\beta^2  \int d^4x_1 d^4x_2 \Big( \hat{\pi}_{\delta\eta}(x), \hat{q}^{\delta}(x_1),\hat{q}^{\eta}(x_2) \Big),\nonumber\\
	\eta_{\pi j j} &= \frac{1}{5} \beta \int d^4x_1 d^4x_2 \Big( \hat{\pi}_{\delta\eta}(x), \hat{j}^{\delta}(x_1),\hat{j}^{\eta}(x_2) \Big),\nonumber\\
	\eta_{\pi q j} &= \frac{-1}{5} \beta \int d^4x_1 d^4x_2 \Big( \hat{\pi}_{\delta\eta}(x), \hat{q}^{\delta}(x_1),\hat{j}^{\eta}(x_2) \Big).\nonumber
\end{align}

\subsubsection{Final Expression}

Combining~\eqref{pifirst} and~\eqref{eta_first} 
with~\eqref{finalpi2nl},~\eqref{finalpi2et} 
and~\eqref{finalpi3p}, we get the full expression of the shear stress 
tensor up to second order in gradient expansion, given by
\begin{align}\label{finalpi}
	\big< \hat{\pi}_{\mu\nu}(x) \big> &= 2\eta \sigma_{\mu\nu}+
	2\eta_\omega \left[ \Delta_{\mu\nu\rho\sigma} D\sigma^{\rho\sigma} +\theta \Gamma \sigma_{\mu\nu} \right]\nonumber\\
	&~~~ +\eta_{\pi\pi\pi} \sigma_{\alpha <\mu}\sigma_{\nu>}^{~\alpha} + \eta_{\pi\tau\tau} \xi_{\alpha <\mu}\xi_{\nu>}^{~\alpha}+2\eta_{\pi\pi\tau} \sigma_{\alpha <\mu}\xi_{\nu>}^{~\alpha}+ 2\eta_{\pi\pi p} \sigma_{\mu\nu}\theta \nonumber\\
	&~~~ +\eta_{\pi qq}M_{< \mu}M_{\nu>} +\eta_{\pi jj}\beta^{-1}\nabla_{< \mu}\alpha \nabla_{\nu>}\alpha + 2\eta_{\pi qj} M_{<\mu}\nabla_{\nu>}\alpha.
\end{align}
Using~\eqref{transport}, and approximating $\sigma^{\rho\sigma}\sim \frac{\pi^{\rho\sigma}}{2\eta}$ in the term $D\sigma^{\rho\sigma}$, we get the evolution equation for the shear stress tensor, given by
\begin{align}\label{relaxpi}
	\pi_{\mu\nu}+\tau_\pi \dot{\pi}_{\mu\nu}&= 2\eta \sigma_{\mu\nu}
	+\tilde{\eta}_\tau \theta \pi_{\mu\nu} +2\eta_\omega \Gamma\sigma_{\mu\nu}\theta \nonumber\\
	&~~~ + 2\eta_{\pi\pi p} \sigma_{\mu\nu}\theta +\eta_{\pi\pi\pi} \sigma_{\alpha <\mu}\sigma_{\nu>}^{~\alpha} + \eta_{\pi\tau\tau} \xi_{\alpha <\mu}\xi_{\nu>}^{~\alpha}+2\eta_{\pi\pi\tau} \sigma_{\alpha <\mu}\xi_{\nu>}^{~\alpha} \nonumber\\
	&~~~ +\eta_{\pi qq}M_{< \mu}M_{\nu>} +\eta_{\pi jj}\beta^{-1}\nabla_{< \mu}\alpha \nabla_{\nu>}\alpha + 2\eta_{\pi qj} M_{<\mu}\nabla_{\nu>}\alpha.
\end{align}
where
\begin{align}
	\tau_\pi&=-\eta_\omega \eta^{-1}\\
	\tilde{\eta}_\tau&=\tau_\pi \eta^{-1}\beta \left( \Gamma\frac{\partial \eta}{\partial \beta}-\delta\frac{\partial \eta}{\partial \alpha} \right)
\end{align}

\subsection{Bulk Viscous Pressure}

From Appendix~\ref{bulk} Eq.~\eqref{bulk_exp_complete}, we have
\begin{equation}\label{bulk_start}
	\Pi = \big< \hat{p}^* \big>_1+\big< \hat{p}^* \big>_2+ \frac{1}{2}\left(\big< \hat{\varepsilon} \big>_1\right)^2 \frac{\partial^2 p}{\partial \varepsilon^2}+ \frac{1}{2}\left(\big< \hat{n} \big>_1\right)^2 \frac{\partial^2 p}{\partial n^2}+\big< \hat{\varepsilon} \big>_1 \big< \hat{n} \big>_1 \frac{\partial^2 p}{\partial n \partial \varepsilon} -\mathcal{K}_{\mu\nu} \big< \hat{S}^{\mu\nu} \big>_1.
\end{equation}
The first-order corrections to energy density, particle density 
and spin density are given by
\begin{align}
	\big< \hat{\varepsilon} \big>_1 &= -\zeta_{{}_{\varepsilon p}} \theta, \label{ep}\\
	\big< \hat{n}\big>_1 &= -\zeta_{{}_{np}} \theta, \label{np}\\
	\big< \hat{S}^{\mu\nu} \big>_1 &= \zeta_{{}_{S\tau}} \xi^{\mu\nu}.\label{st}
\end{align}
Here we used~\eqref{nlc} and~\eqref{CF} and applied the Curie's 
principle to simplify Eqs.~\eqref{ep},~\eqref{np} 
and~\eqref{st}. The transport coefficients 
$\zeta_{{}_{\varepsilon p}}$, $\zeta_{{}_{n p}}$ and 
$\zeta_{{}_{S\tau}}$ are defined by 
\begin{align}
	\zeta_{{}_{\varepsilon p}} &= \beta \int d^4 x_1 \Big( \hat{\varepsilon}(x), \hat{p}^*(x_1) \Big),\\
	\zeta_{{}_{n p}} &= \beta \int d^4 x_1 \Big( \hat{n}(x), \hat{p}^*(x_1) \Big),\\
	\zeta_{{}_{S\tau}} &= \frac{1}{3} \beta \int d^4 x_1 \Big( \hat{S}^{\lambda\sigma}(x), \hat{\tau}_{\lambda\sigma}(x_1) \Big).
\end{align} 
Using~\eqref{ep},~\eqref{np} and~\eqref{st}, we can 
rewrite~\eqref{bulk_start} as
\begin{equation}
	\Pi = -\zeta\theta + \left[\frac{1}{2}\zeta^2_{{}_{\varepsilon p}} \frac{\partial^2 p}{\partial \varepsilon^2}+ \frac{1}{2}\zeta^2_{{}_{n p}} \frac{\partial^2 p}{\partial n^2}+\zeta_{{}_{\varepsilon p}} \zeta_{{}_{n p}}  \frac{\partial^2 p}{\partial n \partial \varepsilon}\right]\theta^2 -\zeta_{{}_{S\tau}}\mathcal{K}_{\mu\nu} \xi^{\mu\nu} + \big< \hat{p}^* \big>_2 ,\label{bulk_exp2}
\end{equation}
where we used $\big< \hat{p}^* \big>_1=-\zeta\theta$, by 
employing~\eqref{zeta_first} and~\eqref{pfirst}. To complete the 
calculation of bulk viscous pressure, we need to evaluate $\big< 
\hat{p}^* \big>_2$.

\subsubsection{Nonlocal correction from the two-point correlation function: $\big< \hat{p}^* (x)\big>_2^{(2),NL}$}

Invoking Curie's principle and using~\eqref{nlc} 
and~\eqref{CF}, we have
\begin{equation}\label{pnl}
	\big< \hat{p}^*(x) \big>_2^{(2),NL} = -\int d^4 x_1 \Big( \hat{p}^*(x), \hat{p}^*(x_1)\Big)\beta(x_1)\theta(x_1)-\big< \hat{p}^*(x) \big>_1.
\end{equation}
As we have calculated the other nonlocal correction terms, we 
write  $\hat{p}^*$ in terms of the energy-momentum tensor and the 
particle current vector, then proceed by expanding all the 
hydrodynamic quantities around $x_1=x$. 
From~\eqref{totalpressure} and~\eqref{projectionsTN}, we have,
\begin{equation}
	\hat{p}^*(x_1)=-\frac{1}{3}\Delta_{\mu\nu}(x_1)\hat{T}^{\mu\nu}(x_1)-\Gamma(x_1)u_\mu(x_1)u_\nu(x_1)\hat{T}^{\mu\nu}(x_1)-\delta(x_1)u_\mu(x_1) \hat{N}^\mu(x_1).
\end{equation}
Expanding all the hydrodynamic quantities (not operators) around 
$x_1=x$, and neglecting the second-order and higher-order terms, we have
\begin{equation}\label{pinprojection}
	\hat{p}^*(x_1)=\left[\hat{p}^*(x_1)\right]_x + (x_1-x)^\lambda \left[\frac{\partial \hat{p}^*(x_1) }{\partial x_1^\lambda}\right]_x ,
\end{equation}
where,
\begin{align}
	\left[\hat{p}^*(x_1)\right]_x &= -\frac{1}{3}\Delta_{\mu\nu}(x)\hat{T}^{\mu\nu}(x_1)-\Gamma(x)u_\mu(x)u_\nu(x)\hat{T}^{\mu\nu}(x_1)-\delta(x)u_\mu(x) \hat{N}^\mu(x_1),\\
	\left[\frac{\partial \hat{p}^*(x_1) }{\partial x_1^\lambda}\right]_x &= -\hat{j}^\mu(x_1)\delta(x)  \frac{\partial u_\mu(x_1) }{\partial x_1^\lambda}\bigg|_x - \hat{\varepsilon}(x_1) \frac{\partial \Gamma(x_1) }{\partial x_1^\lambda}\bigg|_x -\hat{n}(x_1) \frac{\partial \delta(x_1)}{\partial x_1^\lambda}\bigg|_x.  \label{delp_def}
\end{align}
Since we are working in Landau frame, we have set $\hat{h}^\mu=0$ 
while calculating~\eqref{pinprojection}. We can see that there is 
no contribution from the antisymmetric counterpart of heat flow 
vector, {\it i.e.} boost heat vector $\hat{q}^\mu$. 
Substituting~\eqref{pinprojection} in~\eqref{pnl} and expanding 
$\beta\theta$ around $x_1=x$, we get
\begin{align}
	\big< \hat{p}^*(x) \big>_2^{(2),NL} &= -\int d^4 x_1 \left( \hat{p}^*(x), \left[\hat{p}^*(x_1)\right]_x + (x_1-x)^\lambda \left[\frac{\partial \hat{p}^*(x_1) }{\partial x_1^\lambda}\right]_x \right)\Bigg[\beta(x)\theta(x) + (x_1-x)^\lambda\frac{\partial(\beta\theta)}{\partial x_1^\lambda}\bigg|_{x}\Bigg]-\big< \hat{p}^*(x) \big>_1\nonumber\\
	&= -\frac{\partial(\beta\theta)}{\partial x_1^\lambda}\bigg|_x \int d^4x_1 \Big( \hat{p}^*(x),\left[\hat{p}^*(x_1)\right]_x \Big)(x_1-x)^\lambda -\beta(x)\theta(x) \int d^4x_1 \left( \hat{p}^*(x),\left[ \frac{\partial \hat{p}^*(x_1)}{\partial x_1^\lambda}\right]_x \right)(x_1-x)^\lambda, \label{p_term2}
\end{align}
where the leading-order term cancels $\big< \hat{p}^*(x) \big>_1$ 
using~\eqref{pfirst}. Further, we approximate the first term 
in~\eqref{p_term2} with 
$\left[\hat{p}^*(x_1)\right]_x=\hat{p}^*(x_1)$,  because it 
already has a second-order term. Also, we use~\eqref{delp_def} to 
solve the second term in~\eqref{p_term2}. Thus, we obtain
\begin{align}
	\big< \hat{p}^* \big>_2^{(2),NL} &= -\partial_\lambda(\beta\theta)\beta^{-1} K^\lambda_p +\theta (\partial_\lambda \Gamma)  K^\lambda_\varepsilon +\theta(\partial_\lambda \delta) K^\lambda_n, \label{p2step4}
\end{align}
where we have omitted the argument $x$ from all quantities. Also, 
we defined
\begin{align}
	K^\lambda_p &=\beta(x)\int d^4x_1 \Big( \hat{p}^*(x),\hat{p}^*(x_1) \Big)(x_1-x)^\lambda =\zeta^\omega_{pp}u^\lambda,\label{Kp}\\
	K^\lambda_\varepsilon &=\beta(x)\int d^4x_1 \Big( \hat{p}^*(x),\hat{\varepsilon}(x_1) \Big)(x_1-x)^\lambda =\zeta^\omega_{p\varepsilon}u^\lambda,\label{Ke}\\
	K^\lambda_n &=\beta(x)\int d^4x_1 \Big( \hat{p}^*(x),\hat{n}(x_1) \Big)(x_1-x)^\lambda =\zeta^\omega_{pn}u^\lambda, \label{Kn}
\end{align}
where $\zeta^\omega_{pp}$, $\zeta^\omega_{p\varepsilon}$, and 
$\zeta^\omega_{pn}$ are related to the frequency-dependent transport 
coefficients (see Eq.~\eqref{freq_dep_trans} in 
Appendix~\ref{correlators}). 
Substituting,~\eqref{Kp},~\eqref{Ke}, and~~\eqref{Kn} 
in~\eqref{p2step4}, we find
\begin{align}
	\big< \hat{p}^* \big>_2^{(2),NL} &= -D(\beta\theta)\beta^{-1} \zeta^\omega_{pp} +\theta (D \Gamma)  \zeta^\omega_{p\varepsilon} +\theta(D \delta) \zeta^\omega_{pn} .
\end{align}
We have three independent thermodynamic variables, hence, we can 
calculate the comoving derivatives, $D\Gamma$ and $D\delta$, in a 
similar way as we calculated $D\beta$ in~\eqref{dbeta}. Further, 
we use the value of $D\varepsilon$, $Dn$ and, $DS^{\alpha\beta}$ 
from the equations of motion, up to the first order in gradient. 
Thus, we obtain
\begin{align}
	\big< \hat{p}^*\big>_2^{(2),NL} &= -\zeta^\omega_{pp}D\theta  -\theta^2[\Gamma\zeta^\omega_{pp} +\tilde{\Gamma}\zeta^\omega_{p\varepsilon} +\tilde{\delta} \zeta^\omega_{pn}], \label{final_p_2nl}
\end{align}
where $\tilde{\Gamma}$ and $\tilde{\delta}$ are defined as
\begin{align}
	\tilde{\Gamma} = h\frac{\partial \Gamma}{\partial \varepsilon} + n\frac{\partial \Gamma}{\partial n} + S^{\alpha\beta}\frac{\partial \Gamma}{\partial S^{\alpha\beta}}~ \text{and}~ \tilde{\delta} = h\frac{\partial \delta}{\partial \varepsilon} + n\frac{\partial \delta}{\partial n} + S^{\alpha\beta}\frac{\partial \delta}{\partial S^{\alpha\beta}}.
\end{align}

\subsubsection{Local correction from two-point correlation function using extended thermodynamic forces: $\big< \hat{p}^*(x) \big>_2^{(2),ET}$}

Using the definition~\eqref{etc} with~\eqref{CS} and employing 
Curie's principle, we can write
\begin{equation}
	\big< \hat{p}^*(x) \big>_2^{(2),ET}= \int d^4 x_1 \Big( \hat{p}^*(x), \beta(x_1)\sum_{i=1,2}\Big[\big(\hat{\mathfrak{D}}_i\partial^i_{\varepsilon n}\beta\big)\mathcal{X} +\big(\hat{\mathfrak{D}}_i\partial^i_{\varepsilon n}\alpha\big)\mathcal{Y} + \big(\hat{\mathfrak{D}}_i\partial^i_{\varepsilon n}\Omega_{\mu\nu}\big) \mathcal{Z}^{\mu\nu}\Big] \Big)\Big|_{local}.\label{q2_term2}
\end{equation}
Since we have to extract the local terms from~\eqref{q2_term2}, 
which can be achieved by expanding all the hydrodynamic 
quantities around $x_1=x$, we get
\begin{equation}\label{final_p_2et}	
	\big< \hat{p}^*(x) \big>_2^{(2),ET}= \sum_{i=1,2}\zeta_{{}_{p\mathfrak{D}_i}}(x)\Big[ \big(\partial^i_{\varepsilon n}\beta\big)_x\mathcal{X}(x)+\big(\partial^i_{\varepsilon n}\alpha\big)_x\mathcal{Y}(x)+\big(\partial^i_{\varepsilon n}\Omega_{\mu\nu}\big)_x \mathcal{Z}^{\mu\nu}(x)\Big],
\end{equation}
where we defined
\begin{equation}
	\zeta_{{}_{p\mathfrak{D}_i}}(x)=\beta(x)\int d^4 x_1 \Big( \hat{p}^*(x),\hat{\mathfrak{D}}_i(x_1)\Big).
\end{equation}

\subsubsection{Local correction from three-point correlation function: $\big< \hat{p}^*(x) \big>_2^{(3)}$}

From~\eqref{tpc}, we can write
\begin{equation}
	\big< \hat{p}^*(x) \big>_2^{(3)} = \int d^4 x_1 d^4 x_2 \Big( \hat{p}^*(x), \hat{C}_F(x_1),\hat{C}_F(x_2) \Big)\Big|_{local},\label{p_3_def}
\end{equation}
where $\hat{C}_F$ is given by the Eq~\eqref{CF}. The 
nonvanishing correlators are
\begin{align}
	\Big( \hat{p}^*(x), \hat{p}^*(x_1),\hat{p}^*(x_2) \Big) &= \Big( \hat{p}^*(x), \hat{p}^*(x_1),\hat{p}^*(x_2) \Big), \label{p3corr1}\\
	\Big( \hat{p}^*(x), \hat{j}_\mu(x_1),\hat{j}_\nu(x_2) \Big) &=\frac{1}{3}\Delta_{\mu\nu} \Big( \hat{p}^*(x), \hat{j}^\alpha(x_1),\hat{j}_\alpha(x_2) \Big),\label{p3corr2}\\
	\Big( \hat{p}^*(x), \hat{q}_\mu(x_1),\hat{q}_\nu(x_2) \Big) &=\frac{1}{3}\Delta_{\mu\nu} \Big( \hat{p}^*(x), \hat{q}^\alpha(x_1),\hat{q}_\alpha(x_2) \Big),\label{p3corr3}\\
	\Big( \hat{p}^*(x), \hat{\pi}_{\mu\nu}(x_1),\hat{\pi}_{\alpha\beta}(x_2) \Big) &= \frac{1}{5}\Delta_{\mu\nu\alpha\beta} \Big( \hat{p}^*(x), \hat{\pi}^{\lambda\sigma}(x_1),\hat{\pi}_{\lambda\sigma}(x_2) \Big), \label{p3corr4}\\
	\Big( \hat{p}^*(x), \hat{\tau}_{\mu\nu}(x_1),\hat{\tau}_{\alpha\beta}(x_2) \Big) &= \frac{1}{3}\D_{\mu\nu\alpha\beta} \Big( \hat{p}^*(x), \hat{\tau}^{\lambda\sigma}(x_1),\hat{\tau}_{\lambda\sigma}(x_2) \Big).\label{p3corr5}
\end{align}
All other correlators vanish in an isotropic medium. 
Substituting~\eqref{CF} in~\eqref{p_3_def} and 
using~\eqref{p3corr1}-\eqref{p3corr5}, we obtain
\begin{equation}\label{final_p_3p}
	\big< \hat{p}^* \big>_2^{(3)}	= \zeta_{{}_{ppp}}\theta^2 + \zeta_{{}_{pqq}}M_\mu M^\mu + \zeta_{{}_{p\pi\pi}} \sigma_{\mu\nu}\sigma^{\mu\nu} +\zeta_{{}_{p\tau\tau}} \xi_{\mu\nu}\xi^{\mu\nu} +  \beta^{-1}\zeta_{{}_{pjj}} \nabla_\mu \alpha \nabla^\mu \alpha ,
\end{equation}
where, the transport coefficients are defined by
\begin{align}
	\zeta_{{}_{ppp}} &= \beta^2\int d^4 x_1 d^4 x_2 \Big( \hat{p}^*(x), \hat{p}^*(x_1),\hat{p}^*(x_2) \Big), \\
	\zeta_{{}_{pqq}} &= \frac{\beta^2}{3} \int d^4 x_1 d^4 x_2 \Big( \hat{p}^*(x), \hat{q}^\alpha(x_1),\hat{q}_\alpha(x_2) \Big),\\
	\zeta_{{}_{p\pi\pi}} &= \frac{\beta^2}{5} \int d^4 x_1 d^4 x_2\Big( \hat{p}^*(x), \hat{\pi}^{\lambda\sigma}(x_1),\hat{\pi}_{\lambda\sigma}(x_2) \Big),\\
	\zeta_{{}_{p\tau\tau}} &= \frac{\beta^2}{3} \int d^4 x_1 d^4 x_2\Big( \hat{p}^*(x), \hat{\tau}^{\lambda\sigma}(x_1),\hat{\tau}_{\lambda\sigma}(x_2) \Big),\\
	\zeta_{{}_{pjj}} &= \frac{\beta}{3} \int d^4 x_1 d^4 x_2 \Big( \hat{p}^*(x), \hat{j}^\alpha(x_1),\hat{j}_\alpha(x_2) \Big).
\end{align} 

\subsubsection{Final Expression}
Combining~\eqref{final_p_2nl},~\eqref{final_p_2et}, 
and~\eqref{final_p_3p}, we get
\begin{align}
	\big< \hat{p}^* \big>_2 &= -\zeta^\omega_{pp}D\theta  -\theta^2[\Gamma\zeta^\omega_{pp} +\tilde{\Gamma}\zeta^\omega_{p\varepsilon} +\tilde{\delta} \zeta^\omega_{pn}] + \sum_{i=1,2}\zeta_{{}_{p\mathfrak{D}_i}}\Big[ \big(\partial^i_{\varepsilon n}\beta\big)\mathcal{X}+\big(\partial^i_{\varepsilon n}\alpha\big)\mathcal{Y}+\big(\partial^i_{\varepsilon n}\Omega_{\mu\nu}\big) \mathcal{Z}^{\mu\nu}\Big] \nonumber\\
	&~~~+\zeta_{{}_{ppp}}\theta^2 + \zeta_{{}_{pqq}}M_\mu M^\mu + \zeta_{{}_{p\pi\pi}} \sigma_{\mu\nu}\sigma^{\mu\nu} +\zeta_{{}_{p\tau\tau}} \xi_{\mu\nu}\xi^{\mu\nu} +  \beta^{-1}\zeta_{{}_{pjj}} \nabla_\mu \alpha \nabla^\mu \alpha .\label{final_p2}
\end{align}
Substituting~\eqref{final_p2} in~\eqref{bulk_exp2}, we obtain
\begin{align}\label{finalp}
	\Pi &= -\zeta\theta + \left[\frac{1}{2}\zeta^2_{{}_{\varepsilon p}} \frac{\partial^2 p}{\partial \varepsilon^2}+ \frac{1}{2}\zeta^2_{{}_{n p}} \frac{\partial^2 p}{\partial n^2}+\zeta_{{}_{\varepsilon p}} \zeta_{{}_{n p}}  \frac{\partial^2 p}{\partial n \partial \varepsilon}\right]\theta^2 -\zeta_{{}_{S\tau}}\mathcal{K}_{\mu\nu} \xi^{\mu\nu} \nonumber\\
	&~~~ -\zeta^\omega_{pp}D\theta  -\theta^2[\Gamma\zeta^\omega_{pp} +\tilde{\Gamma}\zeta^\omega_{p\varepsilon} +\tilde{\delta} \zeta^\omega_{pn}] + \sum_{i=1,2}\zeta_{{}_{p\mathfrak{D}_i}}\Big[ \big(\partial^i_{\varepsilon n}\beta\big)\mathcal{X}+\big(\partial^i_{\varepsilon n}\alpha\big)\mathcal{Y}+\big(\partial^i_{\varepsilon n}\Omega_{\mu\nu}\big) \mathcal{Z}^{\mu\nu}\Big] \nonumber\\
	&~~~+\zeta_{{}_{ppp}}\theta^2 + \zeta_{{}_{pqq}}M_\mu M^\mu + \zeta_{{}_{p\pi\pi}} \sigma_{\mu\nu}\sigma^{\mu\nu} +\zeta_{{}_{p\tau\tau}} \xi_{\mu\nu}\xi^{\mu\nu} +  \beta^{-1}\zeta_{{}_{pjj}} \nabla_\mu \alpha \nabla^\mu \alpha .
\end{align}
To get the evolution equation for the bulk viscous pressure, we approximate $\theta\sim -\zeta^{-1}\Pi$, in the term 
$\zeta^\omega_{pp}D\theta $. Thus, we obtain
\begin{align}\label{relaxp}
	\Pi +\tau_\Pi \dot{\Pi} &= -\zeta\theta +\tilde{\zeta}_\tau\Pi \theta + \left[\frac{1}{2}\zeta^2_{{}_{\varepsilon p}} \frac{\partial^2 p}{\partial \varepsilon^2}+ \frac{1}{2}\zeta^2_{{}_{n p}} \frac{\partial^2 p}{\partial n^2}+\zeta_{{}_{\varepsilon p}} \zeta_{{}_{n p}}  \frac{\partial^2 p}{\partial n \partial \varepsilon}\right]\theta^2 -\zeta_{{}_{S\tau}}\mathcal{K}_{\mu\nu} \xi^{\mu\nu} \nonumber\\
	&~~~   -\theta^2[\Gamma\zeta^\omega_{pp} +\tilde{\Gamma}\zeta^\omega_{p\varepsilon} +\tilde{\delta} \zeta^\omega_{pn}] + \sum_{i=1,2}\zeta_{{}_{p\mathfrak{D}_i}}\Big[ \big(\partial^i_{\varepsilon n}\beta\big)\mathcal{X}+\big(\partial^i_{\varepsilon n}\alpha\big)\mathcal{Y}+\big(\partial^i_{\varepsilon n}\Omega_{\mu\nu}\big) \mathcal{Z}^{\mu\nu}\Big] \nonumber\\
	&~~~+\zeta_{{}_{ppp}}\theta^2 + \zeta_{{}_{pqq}}M_\mu M^\mu + \zeta_{{}_{p\pi\pi}} \sigma_{\mu\nu}\sigma^{\mu\nu} +\zeta_{{}_{p\tau\tau}} \xi_{\mu\nu}\xi^{\mu\nu} +  \beta^{-1}\zeta_{{}_{pjj}} \nabla_\mu \alpha \nabla^\mu \alpha .
\end{align}
where,
\begin{align}
	\tau_\Pi&=-\zeta^\omega_{pp} \zeta^{-1}\\
	\tilde{\zeta}_\tau&=\tau_\Pi \zeta^{-1}\beta \left( \Gamma\frac{\partial \zeta}{\partial \beta}-\delta\frac{\partial \zeta}{\partial \alpha} \right)
\end{align}

\section{Discussion}

The evolution equations for the rotational stress tensor 
($\tau_{\mu\nu}$), shear stress tensor ($\pi_{\mu\nu}$), boost 
heat vector ($q_\mu$) and bulk viscous pressure ($\Pi$) are given 
in~\eqref{relaxtau},~\eqref{relaxpi},~\eqref{relaxq} 
and~\eqref{relaxp}, respectively, which get reduced for
the chargeless case,
\begin{align}
	\tau_\tau \dot{\tau}_{\mu\nu}+\tau_{\mu\nu}
		&= 2\gamma \xi_{\mu\nu} +\tilde{\gamma}_\tau  \tau_{\mu\nu}\theta +2\gamma_\omega \Gamma \xi_{\mu\nu}\theta  +  \gamma_{{}_{\tau S}} \mathcal{K}_{\mu\nu} \theta  \nonumber\\
		&~~~ +2\gamma_{\tau\tau p}\xi_{\mu\nu}\theta+\gamma_{\tau\tau\tau}\xi_{\alpha\cll \mu }\xi_{\nu \crr}^{~\alpha}  +\gamma_{\tau\pi\pi}\sigma_{\alpha\cll \mu }\sigma_{\nu \crr}^{~\alpha} +2\gamma_{\tau\pi\tau}\sigma_{\alpha\cll \mu }\xi_{\nu \crr}^{~\alpha} +\gamma_{\tau qq}M_{\cll \mu}M_{\nu\crr}, \label{c-less-tau}\\
	\tau_\pi \dot{\pi}_{\mu\nu}+\pi_{\mu\nu}&= 2\eta \sigma_{\mu\nu}+\tilde{\eta}_\tau \theta \pi_{\mu\nu} +2\eta_\omega \Gamma \sigma_{\mu\nu}\theta \nonumber\\
		&~~~ + 2\eta_{\pi\pi p} \sigma_{\mu\nu}\theta+\eta_{\pi\pi\pi} \sigma_{\alpha <\mu}\sigma_{\nu>}^{~\alpha} + \eta_{\pi\tau\tau} \xi_{\alpha <\mu}\xi_{\nu>}^{~\alpha}+2\eta_{\pi\pi\tau} \sigma_{\alpha <\mu}\xi_{\nu>}^{~\alpha}  +\eta_{\pi qq}M_{< \mu}M_{\nu>},\label{c-less-pi}\\
	\tau_q \dot{q}_\mu+q_\mu &= -\lambda M_\mu +\tilde{\lambda}_\tau \theta q_\mu+\lambda_\omega \Gamma M_\mu\theta  -\lambda \mathcal{Q}_\mu\nonumber\\
		&~~~  +\lambda_{qqp} M_\mu\theta+ \lambda_{qq\tau} \xi_{\mu\nu}M^\nu +\lambda_{qq\pi} \sigma_{\mu\nu}M^\nu, \label{c-less-q}\\
        \tau_\Pi \dot{\Pi}+\Pi  &= -\zeta\theta +\tilde{\zeta}_\tau\Pi \theta -\theta^2[\Gamma\zeta^\omega_{pp} +\tilde{\Gamma}\zeta^\omega_{p\varepsilon}]+ \frac{1}{2}\zeta^2_{{}_{\varepsilon p}} \frac{\partial^2 p}{\partial \varepsilon^2}\theta^2     -\zeta_{{}_{S\tau}}\mathcal{K}_{\mu\nu} \xi^{\mu\nu}+ \zeta_{{}_{p\mathfrak{D}_1}}\Big[ \big(\partial^1_{\varepsilon n}\beta\big)\mathcal{X}+\big(\partial^1_{\varepsilon n}\Omega_{\mu\nu}\big) \mathcal{Z}^{\mu\nu}\Big]  \nonumber\\
		&~~~+\zeta_{{}_{ppp}}\theta^2 + \zeta_{{}_{pqq}}M_\mu M^\mu + \zeta_{{}_{p\pi\pi}} \sigma_{\mu\nu}\sigma^{\mu\nu} +\zeta_{{}_{p\tau\tau}} \xi_{\mu\nu}\xi^{\mu\nu}.\label{c-less-p}
\end{align}
The second and third terms on the right-hand side in the first line of each evolution equation arise due to the nonlocal 
correction. That is why they contain a finite relaxation time. 
All the terms written in the last line of each evolution equation 
originate from the second-order corrections due to the local 
contribution from the three-point correlation function. They 
represent all possible combinations of different thermodynamic 
forces ($\sigma_{\mu\nu}$, $\xi_{\mu\nu}$, $\theta$ and $M_\mu$) 
present in the system. Transport coefficients coupled to these 
terms are given by the three-point correlation function. 
Comparing to the previous study~\cite{HARUTYUNYAN2022168755}, the 
extra terms, $\eta_{\pi\tau\tau} \xi_{\alpha 
	<\mu}\xi_{\nu>}^{~\alpha}$, $2\eta_{\pi\pi\tau} 
\sigma_{\alpha <\mu}\xi_{\nu>}^{~\alpha}$ and  $\eta_{\pi qq}M_{< 
	\mu}M_{\nu>}$, appearing in the evolution of the shear stress 
tensor~\eqref{c-less-pi} are the artifacts of the new 
thermodynamic forces ($\xi_{\mu\nu}$ and $M_\mu$) that arise due 
to the addition of antisymmetric terms in the energy-momentum tensor.

Additionally, we find that the structure of the evolution 
equations for the shear stress tensor and the rotational stress 
tensor are similar, except for an extra term, $\gamma_{{}_{\tau 
		S}} \mathcal{K}_{\mu\nu} \theta$, which appears in the 
	evolution equation of the rotational stress tensor. This term 
	arises from corrections due to the extended thermodynamic 
	forces via the two-point correlation function. Since we 
	consider the spin density as an independent thermodynamic 
	variable, which is an antisymmetric tensor of rank two, it 
	generates the possibility of a two-point correlation in the 
	presence of the rotational stress tensor—something that is 
	absent in the case of the shear stress tensor.

Moreover, we identify a term, 
$\zeta_{{}_{S\tau}}\mathcal{K}_{\mu\nu} \xi^{\mu\nu}$, involving 
the transport coefficient, given by the correlation of spin 
density with the rotational stress tensor, in the evolution of 
bulk viscous pressure. The origin of this term can be traced to 
the fact that the system now includes an additional independent 
thermodynamic variable, namely the spin density. Consequently, 
the equilibrium pressure depends on the spin density. Therefore, 
any deviation from equilibrium, which is nothing but the bulk 
viscous pressure, must also depend on the spin density.

Apart from that, within the framework of our current ordering scheme, we observe that the spin density operator ($\hat{S}^{\mu\nu}$) correlates only with the rotational stress tensor ($\hat{\tau}^{\mu\nu}$) through the two-point correlation function. This correlation gives rise to two novel transport coefficients, $\gamma_{{}_{\tau S}}$ and $\zeta_{{}_{S\tau}}$. These coefficients, which emerge at the second order, encapsulate the interplay between the antisymmetric component of the energy-momentum tensor and the spin density tensor. Their existence can be attributed to the inclusion of spin density as an independent thermodynamic variable in the system. The introduction of these transport coefficients is significant, as they provide critical insights into the role and influence of spin density on the medium. Their study is essential for understanding the spin-induced transport properties of the system.

Further, we compare our results for the shear-stress tensor and 
the bulk viscous pressure with the results obtained for 
nonconformal fluids in~\cite{Romatschke:2009kr}. The most 
general second-order expressions for the shear-stress tensor and 
bulk viscous pressure of a nonconformal fluid at zero particle 
chemical potential ($\mu=0$) and zero spin chemical potential 
($\omega_{\alpha\beta}=0$) in flat space-time are given by
\begin{align}
	\pi_{\mu\nu}&=2\eta \sigma_{\mu\nu} -2\eta\tau_{\pi}\left[ \Delta_{\mu\nu\alpha\beta}D\sigma^{\alpha\beta} +\frac{\theta}{3}\sigma_{\mu\nu} \right] -2\eta \tau^*_{\pi}\frac{\theta}{3}\sigma_{\mu\nu} \nonumber\\
	&~~~ +\lambda_1 \sigma_{\alpha <\mu}\sigma_{\nu>}^{~\alpha} +\lambda_2 \sigma_{\alpha <\mu}\overline{\omega}_{\nu>}^{~\alpha} +\lambda_3 \overline{\omega}_{\alpha <\mu}\overline{\omega}_{\nu>}^{~\alpha} +\lambda_4 \nabla_{< \mu}\ln s \nabla_{\nu>}\ln s, \label{non-con-pi}\\
	\Pi &= -\zeta \theta + \zeta \tau_{\Pi} D\theta + \xi_2 \theta^2 +\xi_1 \sigma_{\mu\nu}\sigma^{\mu\nu}+ \xi_3 \overline{\omega}_{\mu\nu}\overline{\omega}^{\mu\nu} + \xi_4 \nabla_\mu \ln s \nabla^\mu \ln s ,\label{non-con-p}
\end{align}
respectively. Here, 
$\overline{\omega}_{\mu\nu}=\D_{\mu\nu}^{\alpha\beta}\partial_\alpha u_\beta$ is the vorticity tensor. 

Next, we rewrite our results for the shear stress 
tensor~\eqref{finalpi} and the bulk viscous 
pressure~\eqref{finalp} for the case of a chargeless and spinless 
fluid with particle chemical potential $\mu=0$ and spin chemical 
potential $\omega_{\mu\nu}=0$. Thus we obtain
\begin{align}
	\pi_{\mu\nu} &= 2\eta \sigma_{\mu\nu}+
	2\eta_\omega \left[ \Delta_{\mu\nu\rho\sigma} D\sigma^{\rho\sigma} +\theta \Gamma \sigma_{\mu\nu} \right]+ 2\eta_{\pi\pi p} \sigma_{\mu\nu}\theta\nonumber\\
	&~~~ +\eta_{\pi\pi\pi} \sigma_{\alpha <\mu}\sigma_{\nu>}^{~\alpha} +2\eta_{\pi\pi\tau} \sigma_{\alpha <\mu}\overline{\omega}_{\nu>}^{~\alpha}+ \eta_{\pi\tau\tau} \overline{\omega}_{\alpha <\mu}\overline{\omega}_{\nu>}^{~\alpha} +\eta_{\pi qq}M_{< \mu}M_{\nu>} \label{reduced-pi}\\
	\Pi &= -\zeta\theta -\zeta^\omega_{pp}D\theta + \left[\zeta_{{}_\theta}+\zeta_{{}_{ppp}}\right]\theta^2 + \zeta_{{}_{p\pi\pi}} \sigma_{\mu\nu}\sigma^{\mu\nu} +\zeta_{{}_{p\tau\tau}} \overline{\omega}_{\mu\nu}\overline{\omega}^{\mu\nu} + \zeta_{{}_{pqq}}M_\mu M^\mu+ \zeta_{{}_{p\varepsilon}} \frac{\partial\beta}{\partial \varepsilon}\mathcal{X} \label{reduced-p}.
\end{align}
where we defined 
$\zeta_{{}_\theta}=\frac{1}{2}\zeta^2_{{}_{\varepsilon p}} 
\frac{\partial^2 p}{\partial \varepsilon^2}-\Gamma\zeta^\omega_{pp} -\tilde{\Gamma}\zeta^\omega_{p\varepsilon}$. Also, we want to 
emphasize on the point that, for a chargeless and spinless case, 
$M_\mu$ can be written as 
$M_\mu=\frac{2(\varepsilon+p)}{\beta}\frac{\partial\beta}{\partial\varepsilon}\nabla_\mu \ln s$, by employing the ideal 
hydrodynamic equation $(\varepsilon+p)Du_\mu=\nabla_\mu p$ 
together with the thermodynamics relations, $Tds=d\varepsilon$, $dp=sdT$ and $\varepsilon+p=Ts$. 

Now, It can be seen that Eq.~\eqref{reduced-pi} produces 
all the terms contained in~\eqref{non-con-pi}, and the 
corresponding transport coefficients in each equation can be 
compared. However, for the case of bulk viscous 
pressure~\eqref{reduced-p}, there is an additional term 
$\zeta_{{}_{p\varepsilon}} \frac{\partial\beta}{\partial \varepsilon}\mathcal{X}$ which originats from the two-point 
correlation function using the extended thermodynamic forces.

\section{Summary and outlook}

We have derived the second-order expressions for the dissipative 
currents $\tau_{\mu\nu}$, $q_\mu$, $\pi_{\mu\nu}$, and $\Pi$ 
using Zubarev's nonequilibrium statistical operator formalism, 
incorporating the full energy-momentum tensor up to the first 
order, including both symmetric and antisymmetric terms. 
Previously, the second-order expressions for $\pi_{\mu\nu}$ and 
$\Pi$ were obtained using only the symmetric energy-momentum 
tensor \cite{HARUTYUNYAN2022168755}. Our calculations extend 
these second-order expressions for $\pi_{\mu\nu}$ and $\Pi$, 
introducing additional terms that account for the effects of spin 
degrees of freedom present in the system. Additionally, we have 
treated the spin density as an independent thermodynamic variable 
in our analysis. Our findings indicate that the spin density 
primarily influences $\tau_{\mu\nu}$ and $\Pi$. This influence 
manifests only in the second order, which is attributed to the 
gradient scheme of the spin chemical potential, which is 
considered first-order in the gradient expansion in this work. Furthermore, new transport coefficients arise in the second order, and each transport coefficient is related to a particular correlation function. We have presented a formula for each transport coefficient in terms of a correlation function. Finally, the second-order expressions of each dissipative tensor are used to find the evolution equation of the respective tensors.

 As an outlook, there are a few research directions along the lines of this work. One can calculate the transport coefficients using quantum field theory with the help of Kubo formulas. Recently, work has been performed to calculate the Kubo-formulas, taking the spin chemical potential as a leading order term~\cite{Dey:2024cwo}. Also, one can rederive the second-order spin hydrodynamics using the full antisymmetric spin tensor that naturally includes the pseudo gauge freedom. Recently, we came across \cite{She:2024rnx}, where a second-order spin hydrodynamics theory was investigated using a generalized form of the spin tensor. Furthermore, it would be interesting to see the cross-effect in second order by enforcing a more general positivity condition on entropy production~\cite{PhysRevC.107.024915}. Studying the second-order spin hydrodynamics and the evolution of dissipative tensors, combined with the numerical simulation, can give a better understanding of the evolution of the relativistic medium containing spin degrees of freedom.

\section*{Acknowledgment}
We would like to thank Dirk H. Rischke for some
important clarification in choosing spin density as an
independent thermodynamic variable. We would also like
to thank Amaresh Jaiswal and Koichi Hattori for clearing a
few doubts on spin hydrodynamics. A. T. further thanks
Arpan Das and Amaresh Jaiswal for their valuable input
during the conceptual discussions for the comments on
this paper.

%


\begin{appendix}
\section{Projection Tensors}\label{tensor}
\subsection{Fourth rank projection tensors}
The two fourth-rank projection tensor used in this work are 
the symmetric traceless projector ($\Delta_{\mu\nu\alpha\beta}$), and 
antisymmetric projector ($\D_{\mu\nu\alpha\beta}$). They both 
are orthogonal to fluid velocity, and defined by
\begin{align}
	\Delta_{\mu\nu\alpha\beta} &= \frac{1}{2} \left( \Delta_{\mu\alpha}\Delta_{\nu\beta} + \Delta_{\mu\beta}\Delta_{\nu\alpha} \right) -\frac{1}{3}\Delta_{\mu\nu}\Delta_{\alpha\beta}, \\
	\D_{\mu\nu\alpha\beta} &= \frac{1}{2} \left( \Delta_{\mu\alpha}\Delta_{\nu\beta} - \Delta_{\mu\beta}\Delta_{\nu\alpha} \right).
\end{align}
Their properties are 
\begin{align}\label{properties_projectors}
	& u^\mu\Delta_{\mu\nu\alpha\beta}=0,~~ \Delta_{\mu\nu\alpha\beta} =\Delta_{\nu\mu\alpha\beta}=\Delta_{\alpha\beta\mu\nu},~~ \Delta^\mu_\rho \Delta_{\mu\nu\alpha\beta} = \Delta_{\rho\nu\alpha\beta},\nonumber\\
	& \Delta_{\mu\nu\alpha\beta}\Delta^{\alpha\beta}_{\rho\sigma}=\Delta_{\mu\nu\rho\sigma},~~ \Delta_{\mu\h\alpha\beta}^{\h\mu}=0,~~ \Delta_{\mu\nu\h\beta}^{\h\h\nu}=\frac{5}{3}\Delta_{\mu\beta},~~ \Delta_{\mu\nu}^{\h\h\mu\nu}=5,\nonumber\\
	& u^\mu\D_{\mu\nu\alpha\beta}=0,~~ \D_{\mu\nu\alpha\beta} =-\D_{\nu\mu\alpha\beta}=\D_{\alpha\beta\mu\nu},~~ \Delta^\mu_\rho \D_{\mu\nu\alpha\beta} = \D_{\rho\nu\alpha\beta},\nonumber\\
	&\D_{\mu\nu\alpha\beta}\D^{\alpha\beta}_{\rho\sigma}=\D_{\mu\nu\rho\sigma},~~\D_{\mu\h\alpha\beta}^{\h\mu}=0,~~\D_{\mu\nu\h\beta}^{\h\h\nu}=-\Delta_{\mu\beta},~~\D_{\mu\nu}^{\h\h\mu\nu}=3.
\end{align}

\subsection{Nonlocal Projectors}\label{non-local_projectors}
The nonlocal projectors are defined by
\begin{align}
	\Delta_{\mu\nu}(x,y)&=\Delta_{\mu\lambda}(x)\Delta^\lambda_\nu(y),\label{nondelta_def}\\
	\Delta_{\mu\nu\alpha\beta}(x,y)&=\Delta_{\mu\nu\rho\sigma}(x)\Delta^{\rho\sigma}_{\alpha\beta}(y),\label{nonsydelta_def}\\
	\D_{\mu\nu\alpha\beta}(x,y)&=\D_{\mu\nu\rho\sigma}(x)\D^{\rho\sigma}_{\alpha\beta}(y),\label{nonasdelta_def}
\end{align}
with some straightforward orthogonal properties, given by 
\begin{align}\label{properties_nonlocalprojectors}
	u^\mu(x)\Delta_{\mu\nu}(x,y)&=\Delta_{\mu\nu}(x,y)u^\nu(y)=0, \\
	u^\mu(x)\Delta_{\mu\nu\alpha\beta}(x,y)=u^\nu(x)\Delta_{\mu\nu\alpha\beta}(x,y)&=\Delta_{\mu\nu\alpha\beta}(x,y)u^\alpha(y)=\Delta_{\mu\nu\alpha\beta}(x,y)u^\beta(y)=0, \\
	u^\mu(x)\D_{\mu\nu\alpha\beta}(x,y)=u^\nu(x)\D_{\mu\nu\alpha\beta}(x,y)&=\D_{\mu\nu\alpha\beta}(x,y)u^\alpha(y)=\D_{\mu\nu\alpha\beta}(x,y)u^\beta(y)=0 .
\end{align}
In order to write a hydrodynamic gradient expansion for the 
projectors, we expand the projectors around a fixed point. We 
start by the Taylor expansion for the fluid velocity and the second-rank projection tensor, given by
\begin{align}
	u_\mu(y)&=u_\mu(x)+(y-x)^\alpha \frac{\partial u_\mu(y)}{\partial y^\alpha}\bigg|_{y=x} + \cdots, \label{fluid_ex} \\
	\Delta_{\mu\nu}(y)&=\Delta_{\mu\nu}(x)-(y-x)^\alpha\left[ u_\mu(x)\frac{\partial u_\nu (y)}{\partial y^\alpha}\bigg|_{y=x}+u_\nu(x)\frac{\partial u_\mu (y)}{\partial y^\alpha}\bigg|_{y=x} \right]+\cdots. \label{secondprojec_ex}
\end{align}
As discussed in the Sec.~\ref{FO_SH}, the second term 
in~\eqref{fluid_ex} and~\eqref{secondprojec_ex} are of first 
order in hydrodynamic gradient expansion. We can conclude that up 
to first order $u_\mu(y)u^\mu(x)\approx1$. With the help 
of~\eqref{fluid_ex} and using the definition of 
$\Delta^\lambda_\nu(y)$, we can rewrite~\eqref{nondelta_def} as
\begin{align}\label{nondelta_ex}
	\Delta_{\mu\nu}(x,y) = \Delta_{\mu\nu}(x)-u_\nu(x) (y-x)^\alpha \frac{\partial u_\mu(y)}{\partial y^\alpha}\bigg|_{y=x} + \cdots.
\end{align}
Now, we can derive some useful properties of nonlocal projectors 
which are approximated up to the first order in gradient 
correction. Using~\eqref{nondelta_ex}, we obtain
\begin{align}\label{approx_properties_nonlocalprojectors}
	\Delta_{\mu\nu}(x,y)\Delta^\nu_\lambda(x)\approx \Delta_{\mu\lambda}(x),~~ \Delta_{\alpha\beta}(x)\Delta^{\alpha\beta}(y)\approx 3.
\end{align} 
Here, these properties are only valid in the first-order gradient 
approximation. In the same approximation, the nonlocal fourth 
rank symmetric traceless projector~\eqref{nonsydelta_def}, can be 
written as
\begin{align}\label{sydelta_ex}
	\Delta_{\mu\nu\alpha\beta}(x,y)\approx\frac{1}{2}\left[ \Delta_{\mu\alpha}(x,y)\Delta_{\nu\beta}(x,y)+\Delta_{\mu\beta}(x,y)\Delta_{\nu\alpha}(x,y) \right]-\frac{1}{3}\Delta_{\mu\nu}(x)\Delta_{\alpha\beta}(y),
\end{align}
whereas, the nonlocal fourth-rank antisymmetric 
projector~\eqref{nonasdelta_def}, can be rewritten as
\begin{align}\label{asdelta_ex}
	\D_{\mu\nu\alpha\beta}(x,y)=\frac{1}{2}\left[ \Delta_{\mu\alpha}(x,y)\Delta_{\nu\beta}(x,y)-\Delta_{\mu\beta}(x,y)\Delta_{\nu\alpha}(x,y) \right].
\end{align}
Here, this equation is valid up to all orders in gradient 
expansion.

Using~\eqref{nondelta_ex} in~\eqref{sydelta_ex} 
and~\eqref{asdelta_ex}, we obtain
\begin{align}
	\Delta_{\mu\nu\alpha\beta}(x,y)&=\Delta_{\mu\nu\alpha\beta}(x)-(y-x)^\rho \left[ \Delta_{\mu\nu\alpha\lambda}(x)u_\beta(x)+\Delta_{\mu\nu\lambda\beta}(x)u_\alpha(x) \right] \frac{\partial u^\lambda(y)}{\partial y^\rho}\bigg|_{y=x}+\cdots, \label{sy_derivative}\\
	\D_{\mu\nu\alpha\beta}(x,y)&=\D_{\mu\nu\alpha\beta}(x)-(y-x)^\rho \left[ \D_{\mu\nu\alpha\lambda}(x)u_\beta(x)+\D_{\mu\nu\lambda\beta}(x)u_\alpha(x) \right] \frac{\partial u^\lambda(y)}{\partial y^\rho}\bigg|_{y=x}+\cdots.\label{asy_derivative}
\end{align}


\section{Correlation Function}\label{correlators}
The derivation given in this section is similar as the one presented 
in~\cite{Hosoya:1983id,Huang:2011dc,HARUTYUNYAN2022168755}.
Starting from the~\eqref{twopointdef}, we have
\begin{equation}
	\Big( \hat{X}(\vec{x},t), \hat{Y}(\vec{x}_1,t_1) \Big) = \int_0^1 d\tau \Big< \hat{X}(\vec{x},t) \left[ e^{-\hat{A}\tau} \hat{Y}(\vec{x}_1,t_1)e^{\hat{A}\tau}-\left< e^{-\hat{A}\tau}\hat{Y}(\vec{x}_1,t_1)e^{\hat{A}\tau} \right>_l \right] \Big>_l, \label{corrfun1}
\end{equation}
Inferring to~\eqref{A_euilibrium}, we can conclude that 
$\hat{A}=\beta \hat{H}$, where $\hat{H}$ is the Hamiltonian of a 
rotating system in a grand canonical ensemble. Therefore, we 
obtain the time evolution of any operator in the Heisenberg picture 
in imaginary time as
\begin{equation}
	\hat{Y}(\vec{x},t+i\tau')=e^{-\hat{H}\tau'}\hat{Y}(\vec{x},t)e^{\hat{H}\tau'},
\end{equation}
which leads to the Kubo-Martin-Schwinger relations, given by
\begin{align}
	\left< \hat{Y}(\vec{x},t+i\tau') \right>_l &= \left< \hat{Y}(\vec{x},t) \right>_l, \label{KMS1}\\
	\left< \hat{X}(\vec{x},t)\hat{Y}(\vec{x}_1,t_1+i\beta) \right>_l &= \left< \hat{Y}(\vec{x}_1,t_1) \hat{X}(\vec{x},t)\right>_l.\label{KMS2}
\end{align}
Using~\eqref{KMS1} and~\eqref{KMS2}, in~\eqref{corrfun1} and 
assuming that the correlations vanish at $t\to\infty$, we can 
establish the relation
\begin{equation}
	\Big( \hat{X}(\vec{x},t), \hat{Y}(\vec{x}_1,t_1) \Big) =-\frac{1}{\beta} \int_{-\infty}^{t_1} dt' \mathcal{G}^R_{\hat{X}\hat{Y}}(\vec{x}-\vec{x}_1,t-t'),\label{kubo_to_green}
\end{equation}
where
\begin{eqnarray}
	 \mathcal{G}^R_{\hat{X}\hat{Y}}(\vec{x}-\vec{x}_1,t-t')=-i\theta(t-t')\left< \left[ \hat{X}(\vec{x},t),  \hat{Y}(\vec{x}_1,t') \right] \right>_l.\label{ret_green}
\end{eqnarray}
Now we consider a general form of frequency-dependent transport 
coefficient, involving a two-point correlation function, given by
\begin{eqnarray}
	\mathcal{I}_{\hat{X}\hat{Y}}(\omega) = \beta \int d^3x_1 \int_{-\infty}^t dt_1 e^{\epsilon(t_1-t)} e^{i\omega(t-t_1)}\Big( \hat{X}(\vec{x},t),\hat{Y}(\vec{x}_1,t_1)  \Big).\label{freq_tran}
\end{eqnarray}
The retarded Green's function obtained in~\eqref{ret_green} is 
transitionally invariant both in space as well as in time. We can 
exploit this property to rewrite the integral~\eqref{freq_tran} 
by setting $(\vec{x},t)=(0,0)$. 
Substituting~\eqref{kubo_to_green} in~\eqref{freq_tran}, we obtain
\begin{equation}
	\mathcal{I}_{\hat{X}\hat{Y}}(\omega)=-\int^0_{-\infty} dt_1 e^{(\epsilon-i\omega)t_1} \int_{-\infty}^{t_1}dt'\int d^3 x_1  \mathcal{G}_{\hat{X}\hat{Y}}^R (-\vec{x}_1,-t').\label{green_step1}
\end{equation} 
Considering the Fourier transformation 
\begin{equation}
	\mathcal{G}^R_{\hat{X}\hat{Y}}(\vec{x}_1,t')=\int \frac{d^3k}{(2\pi)^3} \int_{-\infty}^{\infty}\frac{d\omega'}{2\pi} e^{-i(\omega't'-\vec{k}\cdot\vec{x})} \mathcal{G}_{\hat{X}\hat{Y}}^R (\vec{k},\omega'),
\end{equation} 
and using the definition of the Dirac delta function, we can write
\begin{equation}
	\int d^3x \mathcal{G}_{\hat{X}\hat{Y}}^R (-\vec{x},-t') = \lim_{\vec{k}\to 0} \int_{-\infty}^{\infty}\frac{d\omega'}{2\pi}e^{i\omega't'}\mathcal{G}_{\hat{X}\hat{Y}}^R (\vec{k},\omega').\label{integ_space}
\end{equation}
Substituting~\eqref{integ_space} in~\eqref{green_step1} and 
solving the integral over $t'$ by shifting $\omega'\to 
\omega'-i\delta$, where $\delta\to0^+$, we obtain
\begin{equation}
     \mathcal{I}_{\hat{X}\hat{Y}}(\omega)=\lim_{\delta \to 0^+} \frac{i}{\omega+i\epsilon}\oint \frac{d\omega'}{2\pi i}\left[ \frac{1}{\omega'-(\omega+i\epsilon+i\delta)}-\frac{1}{\omega'-i\delta} \right]\mathcal{G}^R_{\hat{X}\hat{Y}}(\omega').\label{final_integ_I}
\end{equation}
Here, $\mathcal{G}^R_{\hat{X}\hat{Y}}(\omega')\equiv 
\lim_{\vec{k}\to 0}\mathcal{G}^R_{\hat{X}\hat{Y}}(\vec{k},\omega')$. In order to 
solve~\eqref{final_integ_I}, we first complexify $\omega'$, and 
use Cauchy's integral formula together with the residue 
theorem. Inferring~\eqref{final_integ_I}, we obtain two poles at 
$\omega'=\omega+i\epsilon+i\delta$ and $\omega'=i\delta$. Also, 
we close the contour in the upper half plane and assume that, at 
infinity, $\mathcal{G}^R_{\hat{X}\hat{Y}}(\omega')$ vanishes 
sufficiently faster than the denominator ($\frac{1}{\omega'}$). 
Thus, we have
\begin{equation}
	 \mathcal{I}_{\hat{X}\hat{Y}}(\omega)=\lim_{\delta \to 0^+} \frac{i}{\omega+i\epsilon} \left[ \mathcal{G}^R_{\hat{X}\hat{Y}}(\omega+i\epsilon+i\delta)-\mathcal{G}^R_{\hat{X}\hat{Y}}(i\delta) \right].
\end{equation}
Now we can take the limits $\delta\to 0^+$ and $\epsilon\to 0^+$, 
with $\epsilon$ being the irreversibility parameter, and which must be 
set to zero at the end of the calculations. Thus, we obtain
\begin{equation}
	 \mathcal{I}_{\hat{X}\hat{Y}}(\omega)= \frac{i}{\omega} \left[ \mathcal{G}^R_{\hat{X}\hat{Y}}(\omega)-\mathcal{G}^R_{\hat{X}\hat{Y}}(0) \right].\label{kubo_green3}
\end{equation}
For $\omega\to0$,~\eqref{kubo_green3} can be simplified to
\begin{equation}
	\mathcal{I}_{\hat{X}\hat{Y}}(0)= i\lim_{\omega\to0}\frac{d}{d\omega} \mathcal{G}^R_{\hat{X}\hat{Y}}(\omega)= i\lim_{\omega\to0}\lim_{\vec{k}\to0}\frac{\partial}{\partial\omega} \mathcal{G}^R_{\hat{X}\hat{Y}}(\vec{k},\omega).\label{final_green_formula1}
\end{equation}
Now, since the operators $\hat{X}$ and $\hat{Y}$ are Hermitian, 
we have
\begin{equation}
	\left< \left[ \hat{X}(\vec{x},t), \hat{Y}(\vec{x}_1,t_1) \right] \right>_l^* = -\left< \left[ \hat{X}(\vec{x},t), \hat{Y}(\vec{x}_1,t_1) \right] \right>_l,
\end{equation}
which indicates that $\left< \left[ \hat{X}(\vec{x},t), 
\hat{Y}(\vec{x}_1,t_1) \right] \right>_l$ is a purely imaginary 
quantity, and by~\eqref{ret_green}, we can conclude that the 
retarded Green's function $\mathcal{G}^R_{\hat{X}\hat{Y}}$ and 
$\Big( \hat{X}(\vec{x},t), \hat{Y}(\vec{x}_1,t_1) \Big)$ are 
purely real quantities. Therefore, from~\eqref{freq_tran}, we get 
to the conclusion that $	\mathcal{I}_{\hat{X}\hat{Y}}(0)$ 
should be a purely real quantity. Thus we can 
rewrite~\eqref{final_green_formula1} as
\begin{equation}
	\mathcal{I}_{\hat{X}\hat{Y}}(0) = \beta \int d^4 x_1 \Big( \hat{X}(x),\hat{Y}(x_1)\Big)=-\lim_{\omega\to 0}\frac{d}{d\omega}{\rm Im}\mathcal{G}^R_{\hat{X}\hat{Y}}(\omega) =-\lim_{\omega\to 0}\lim_{\vec{k}\to 0}\frac{\partial}{\partial\omega}{\rm Im}\mathcal{G}^R_{\hat{X}\hat{Y}}(\vec{k},\omega).
\end{equation}
The other type of dissipative current obtained in this work is of 
the form
\begin{equation}
		\mathcal{J}^\mu_{\hat{X}\hat{Y}}(\omega) = \beta \int d^4x_1  e^{i\omega(t-t_1)}\Big( \hat{X}(x),\hat{Y}(x_1)  \Big)(x_1-x)^\mu.\label{freq_tran_type2}
\end{equation}
The temporal component is given by
\begin{equation}
	\mathcal{J}^0_{\hat{X}\hat{Y}}(\omega) = i\beta \frac{d}{d\omega} \int d^4x_1 e^{i\omega(t-t_1)}\Big( \hat{X}(x),\hat{Y}(x_1) \Big)=i\frac{d}{d\omega}\mathcal{I}_{\hat{X}\hat{Y}}(\omega).\label{temporal_J} 
\end{equation}
The spatial component of 
$\mathcal{J}^\mu_{\hat{X}\hat{Y}}(\omega)$ vanishes in the local 
rest frame, since the correlator $\Big( \hat{X}(x),\hat{Y}(x_1) 
\Big)$ evaluated in the local rest frame depends on 
$|\vec{x}-\vec{x}_1|$, which makes 
$\mathcal{J}^i_{\hat{X}\hat{Y}}(\omega)$ an odd function of 
$\vec{x}-\vec{x}_1$. Furthermore, considering a Taylor expansion 
of $\mathcal{G}^R_{\hat{X}\hat{Y}}(\omega)$ around $\omega=0$, we have
\begin{equation}
	\mathcal{G}^R_{\hat{X}\hat{Y}}(\omega) =\mathcal{G}^R_{\hat{X}\hat{Y}}(0)+\omega \frac{d}{d\omega}\mathcal{G}^R_{\hat{X}\hat{Y}}(\omega)\bigg|_{\omega=0}+\frac{\omega^2}{2!}\frac{d^2}{d\omega^2}\mathcal{G}^R_{\hat{X}\hat{Y}}(\omega)\bigg|_{\omega=0}+\cdots,
\end{equation}
which leads to 
\begin{equation}
	\mathcal{I}_{\hat{X}\hat{Y}}(\omega)=\frac{i}{\omega}\left[\mathcal{G}^R_{\hat{X}\hat{Y}}(\omega)-\mathcal{G}^R_{\hat{X}\hat{Y}}(0)\right]=i\frac{d}{d\omega}\mathcal{G}^R_{\hat{X}\hat{Y}}(\omega)\bigg|_{\omega=0}+\frac{i\omega}{2!}\frac{d^2}{d\omega^2}\mathcal{G}^R_{\hat{X}\hat{Y}}(\omega)\bigg|_{\omega=0}+\cdots . \label{green_expansion}
\end{equation}
Substituting~\eqref{green_expansion} in~\eqref{temporal_J} and 
taking the limit $\omega\to0$, we have
\begin{align}
	\mathcal{J}^0_{\hat{X}\hat{Y}}(0)&=-\frac{1}{2}\lim_{\omega\to0} \frac{d^2}{d\omega^2}\mathcal{G}^R_{\hat{X}\hat{Y}}(\omega)=-\frac{1}{2}\lim_{\omega\to0} \frac{d^2}{d\omega^2}{\rm Re}\mathcal{G}^R_{\hat{X}\hat{Y}}(\omega)=-\frac{1}{2}\lim_{\omega\to0}\lim_{\vec{k}\to0} \frac{\partial^2}{\partial\omega^2}{\rm Re}\mathcal{G}^R_{\hat{X}\hat{Y}}(\vec{k},\omega).
\end{align}
Now, we assume $\mathcal{J}^0_{\hat{X}\hat{Y}}(0)=J$, then we can 
write
\begin{equation}
	\beta\int d^4x_1 \Big( \hat{X}(x),\hat{Y}(x_1)  \Big)(x_1-x)^\mu =\mathcal{J}^\mu_{\hat{X}\hat{Y}}(0)=J u^\mu .\label{freq_dep_trans}
\end{equation}


\section{Correlation Function in terms of projectors and traces}\label{correlations_projectors}

The most general decomposition for a two-point correlation 
function involving two second-rank tensors is given by 
\begin{equation}
	\Big( \hat{X}_{\mu\nu}(x),\hat{Y}_{\alpha\beta}(x_1) \Big) = a_1\Delta_{\mu\nu}\Delta_{\alpha\beta}+ a_2 \Delta_{\mu\alpha}\Delta_{\nu\beta} +a_3 \Delta_{\mu\beta}\Delta_{\nu\alpha}.\label{four_ten_deco}
\end{equation}
A similar decomposition can also be given for a three-point 
correlation function involving one second-rank tensor and two 
vectors
\begin{equation}
	\Big( \hat{X}_{\mu\nu}(x),\hat{Y}_\alpha(x_1),\hat{Z}_\beta(x_2) \Big)=b_1\Delta_{\mu\nu}\Delta_{\alpha\beta}+ b_2 \Delta_{\mu\alpha}\Delta_{\nu\beta} +b_3 \Delta_{\mu\beta}\Delta_{\nu\alpha}.\label{three_ten_deco}
\end{equation}
Here, we present the derivation of~\eqref{asydec}. All the other 
relations, similar to~\eqref{four_ten_deco} 
and~\eqref{three_ten_deco}, can be derived in a similar way. 
Substituting $\hat{X}=\hat{\tau}$ and $\hat{Y}=\hat{\tau}$ 
in~\eqref{four_ten_deco}, we demand that~\eqref{four_ten_deco} 
should be antisymmetric in the exchange of 
$\mu\leftrightarrow\nu$ and $\alpha\leftrightarrow\beta$. We 
obtain
\begin{equation}
	a_1=0,~~a_2=-a_3 =a~\text{(say)}.
\end{equation}
Thus, we have
\begin{equation}
	\Big( \hat{\tau}_{\mu\nu}(x),\hat{\tau}_{\alpha\beta}(x_1) \Big) =  a\Big[ \Delta_{\mu\alpha}\Delta_{\nu\beta} - \Delta_{\mu\beta}\Delta_{\nu\alpha}\Big]= 2a \D_{\mu\nu\alpha\beta},\label{four_ten_deco2}
\end{equation}
where $a$ can be obtained by taking the trace 
of~\eqref{four_ten_deco2},
\begin{eqnarray}
	a=\frac{1}{6}\Big( \hat{\tau}_{\lambda\eta}(x),\hat{\tau}^{\lambda\eta}(x_1) \Big).
\end{eqnarray}

Furthermore, we present the derivation for~\eqref{cor1}. All the 
other three-point correlation functions involving three second-rank tensors, given in~\eqref{cor1}-\eqref{cor4} 
and~\eqref{picor1}-\eqref{picor4}, can be calculated in similar 
way. Using~\eqref{general_six_decom}, we can write,
\begin{align}
	\Big( \hat{\tau}_{\mu\nu}(x), \hat{\tau}_{\rho\sigma}(x_1),\hat{\tau}_{\alpha\beta}(x_2) \Big)  &= a_1 \Delta_{\mu\nu}\Delta_{\rho\alpha}\Delta_{\sigma\beta} +a_2 \Delta_{\mu\nu}\Delta_{\rho\beta}\Delta_{\sigma\alpha} +b_1 \Delta_{\rho\sigma}\Delta_{\mu\alpha}\Delta_{\nu\beta} +b_2 \Delta_{\rho\sigma}\Delta_{\mu\beta}\Delta_{\nu\alpha}\nonumber\\
	&~~~ +c_1 \Delta_{\alpha\beta}\Delta_{\mu\rho}\Delta_{\nu\sigma} +c_2 \Delta_{\alpha\beta}\Delta_{\mu\sigma}\Delta_{\nu\rho} +d \Delta_{\mu\nu}\Delta_{\rho\sigma}\Delta_{\alpha\beta} \nonumber\\
	&~~~ +e_1 \Delta_{\mu\rho}\Delta_{\nu\alpha}\Delta_{\sigma\beta} +e_2 \Delta_{\mu\rho}\Delta_{\nu\beta}\Delta_{\sigma\alpha} +e_3 \Delta_{\mu\sigma}\Delta_{\nu\alpha}\Delta_{\rho\beta} +e_4 \Delta_{\mu\sigma}\Delta_{\nu\beta}\Delta_{\rho\alpha} \nonumber\\
	&~~~ +e_5 \Delta_{\mu\alpha}\Delta_{\nu\rho}\Delta_{\sigma\beta} +e_6 \Delta_{\mu\alpha}\Delta_{\nu\sigma}\Delta_{\rho\beta} +e_7 \Delta_{\mu\beta}\Delta_{\nu\rho}\Delta_{\sigma\alpha} +e_8 \Delta_{\mu\beta}\Delta_{\nu\sigma}\Delta_{\rho\alpha}.\label{three_tau_deco}
\end{align}
This expression should be antisymmetric in the exchange of $\mu 
\leftrightarrow \nu$, $\rho\leftrightarrow\sigma$, and 
$\alpha\leftrightarrow\beta$. This concludes that the coefficients 
of projectors with indices $\mu\nu$, $\rho\sigma$ and 
$\alpha\beta$ must vanish. Thus, we have
\begin{equation}
	a_1=a_2=b_1=b_2=c_1=c_2=d=0.
\end{equation}
Furthermore, exchanging $\alpha\leftrightarrow\beta$ 
in~\eqref{three_tau_deco} and then comparing the coefficients of 
similar projector terms in $\Big( \hat{\tau}_{\mu\nu}(x), 
\hat{\tau}_{\rho\sigma}(x_1),\hat{\tau}_{\beta\alpha}(x_2) \Big) 
$ with $-\Big( \hat{\tau}_{\mu\nu}(x), 
\hat{\tau}_{\rho\sigma}(x_1),\hat{\tau}_{\alpha\beta}(x_2) \Big) $, 
we deduce that
\begin{equation}
	e_1=-e_2,~~e_3=-e_4,~~e_6=-e_8,~~e_5=-e_7.
\end{equation}
Similarly, exchanging $\rho\leftrightarrow\sigma$, we have
\begin{equation}
	e_1=-e_3,~~e_5=-e_6,
\end{equation}
and exchanging $\mu\leftrightarrow\nu$, we have
\begin{equation}
	e_1=-e_5 =e \text{(say)}.
\end{equation}
Thus, we obtain
\begin{align}
	\Big( \hat{\tau}_{\mu\nu}(x), \hat{\tau}_{\rho\sigma}(x_1),\hat{\tau}_{\alpha\beta}(x_2) \Big)  &= e\Big( \Delta_{\mu\rho}\Delta_{\nu\alpha}\Delta_{\sigma\beta} - \Delta_{\mu\rho}\Delta_{\nu\beta}\Delta_{\sigma\alpha} - \Delta_{\mu\sigma}\Delta_{\nu\alpha}\Delta_{\rho\beta} + \Delta_{\mu\sigma}\Delta_{\nu\beta}\Delta_{\rho\alpha} \nonumber\\
	&~~~ - \Delta_{\mu\alpha}\Delta_{\nu\rho}\Delta_{\sigma\beta} + \Delta_{\mu\alpha}\Delta_{\nu\sigma}\Delta_{\rho\beta} + \Delta_{\mu\beta}\Delta_{\nu\rho}\Delta_{\sigma\alpha} - \Delta_{\mu\beta}\Delta_{\nu\sigma}\Delta_{\rho\alpha}\Big)\nonumber\\
	&=2e\Big[ \Delta_{\rho\sigma}\D_{\mu\nu\sigma\beta}-\Delta_{\rho\beta}\D_{\mu\nu\sigma\alpha}-\Delta_{\sigma\alpha}\D_{\mu\nu\rho\beta}+\Delta_{\sigma\beta}\D_{\mu\nu\rho\alpha} \Big]. \label{three_tau_deco3}
\end{align}
In order to calculate the value of $e$, we take the trace and use 
the properties~\eqref{properties_projectors}. Thus, we obtain
\begin{equation}
	e=\frac{1}{6}\Big( \hat{\tau}_\lambda^{\h\delta}(x), \hat{\tau}_\delta^{\h\eta}(x_1),\hat{\tau}_\eta^{\h\lambda}(x_2) \Big).
\end{equation}


\section{Bulk Viscous Pressure}\label{bulk}
The nonequilibrium correction to the pressure is given by
\begin{align}
	\Pi &= \big< \hat{p} \big> -p(\varepsilon, n, S^{\mu\nu}),\nonumber\\
	&= \big< \hat{p} \big>_l +\big< \hat{p} \big>_1+\big< \hat{p} \big>_2 -p(\varepsilon, n, S^{\mu\nu})  .
\end{align}
Here, $p(\varepsilon, n, S^{\mu\nu})$ is the equilibrium pressure 
given by the equation of the state. The average of the pressure 
operator over the local equilibrium can be expanded around the 
equilibrium pressure, mathematically, it can be given by
\begin{align}\label{p1}
	\big< \hat{p} \big>_l
	&=p(\varepsilon, n, S^{\mu\nu}) - \Gamma \Delta \varepsilon - \delta \Delta n - \mathcal{K}_{\mu\nu}\Delta S^{\mu\nu} \nonumber\\
	&~~~ + \frac{1}{2}(\Delta \varepsilon)^2 \frac{\partial^2 p}{\partial \varepsilon^2}+ \frac{1}{2}(\Delta n)^2 \frac{\partial^2 p}{\partial n^2}+ \frac{1}{2} \Delta S^{\mu\nu} \Delta S^{\alpha\beta} \frac{\partial}{\partial S^{\alpha\beta}}\bigg(\frac{\partial p }{\partial S^{\mu\nu}}\bigg)\nonumber\\
	&~~~+\frac{1}{2} \times 2 (\Delta \varepsilon) (\Delta n) \frac{\partial^2 p}{\partial n \partial \varepsilon}+\frac{1}{2} \times 2 (\Delta \varepsilon) (\Delta S^{\mu\nu})\frac{\partial}{\partial \varepsilon}\bigg( \frac{\partial p}{\partial S^{\mu\nu}}\bigg)+\frac{1}{2} \times 2 (\Delta n) (\Delta S^{\mu\nu})\frac{\partial}{\partial n}\bigg( \frac{\partial p}{\partial S^{\mu\nu}}\bigg) + \cdots .
\end{align}
The corrections $\Delta \varepsilon$, $\Delta n$, and $\Delta 
S^{\mu\nu}$ are taken up to the second order. Thus,
\begin{equation}
	\Delta \varepsilon= \big< \hat{\varepsilon} \big>_1+ \big< \hat{\varepsilon} \big>_2, ~~~~\Delta n= \big< \hat{n} \big>_1+ \big< \hat{n} \big>_2 ~~\text{and,}~~~~ \Delta S^{\mu\nu}= \big< \hat{S}^{\mu\nu} \big>_1+ \big< \hat{S}^{\mu\nu} \big>_2 .
\end{equation}
Using these in~\eqref{p1} and neglecting all the third and higher 
order terms, we have
\begin{align}
	\big< \hat{p} \big>_l &= p(\varepsilon, n, S^{\mu\nu}) - \Gamma \big< \hat{\varepsilon} \big>_1 -\Gamma \big< \hat{\varepsilon} \big>_2- \delta \big< \hat{n} \big>_1 - \delta \big< \hat{n} \big>_2  \nonumber\\
	&~~~ + \frac{1}{2}\left(\big< \hat{\varepsilon} \big>_1\right)^2 \frac{\partial^2 p}{\partial \varepsilon^2}+ \frac{1}{2}\left(\big< \hat{n} \big>_1\right)^2 \frac{\partial^2 p}{\partial n^2}+\big< \hat{\varepsilon} \big>_1 \big< \hat{n} \big>_1 \frac{\partial^2 p}{\partial n \partial \varepsilon} -\mathcal{K}_{\mu\nu} \big< \hat{S}^{\mu\nu} \big>_1 \nonumber.
\end{align}
Finally, the expression for the bulk viscous pressure, up to 
second order in hydrodynamic gradient expansion, is given by
\begin{equation}\label{bulk_exp_complete}
	\Pi = \big< \hat{p}^* \big>_1+\big< \hat{p}^* \big>_2+ \frac{1}{2}\left(\big< \hat{\varepsilon} \big>_1\right)^2 \frac{\partial^2 p}{\partial \varepsilon^2}+ \frac{1}{2}\left(\big< \hat{n} \big>_1\right)^2 \frac{\partial^2 p}{\partial n^2}+\big< \hat{\varepsilon} \big>_1 \big< \hat{n} \big>_1 \frac{\partial^2 p}{\partial n \partial \varepsilon} -\mathcal{K}_{\mu\nu} \big< \hat{S}^{\mu\nu} \big>_1.
\end{equation}

\end{appendix}




\end{document}